\documentclass[%
 reprint,
showpacs,preprintnumbers,
 amsmath,amssymb
prd,
]{revtex4-1}

\usepackage[english]{babel}
\usepackage[utf8]{inputenc}
\usepackage{graphicx,graphics}
\usepackage{dcolumn}
\usepackage{bm}
\usepackage{hyperref}
\usepackage{blindtext}
\usepackage{mathrsfs}
\usepackage{pstricks}
\usepackage{color}
\usepackage{slashed}
\usepackage{amsmath}
\usepackage{epsfig}
\usepackage{amsfonts}
\usepackage{amssymb}

\def\sgn{\mathrm{sgn}}
\def\Sgn{\mathrm{Sgn}}
\def\beq{\begin{equation}}
\def\eeq{\end{equation}}
\def\bea{\begin{eqnarray}}
\def\eea{\end{eqnarray}}
\def\nn{\nonumber}
\begin{document}

\title{Spontaneous excitation of an atom in a Kerr spacetime}
\author{G. Menezes}
\email{gabrielmenezes@ufrrj.br}
\affiliation{Grupo de F\'isica Te\'orica e Matem\'atica F\'isica, Departamento de F\'isica, Universidade Federal Rural do Rio de Janeiro, 23897-000 Serop\'edica, RJ, Brazil}
%

\begin{abstract}
We consider radiative processes of an atom in a rotating black-hole background. We assume the atom, represented by a hypothetical two-level system, is coupled via a monopole interaction with a massless quantum scalar field prepared in each one of the usual physical vacuum states of interest. We constrain ourselves to two different states of motion for the atom, namely a static situation in which the atom is placed at a fixed radial distance, and also the case in which it has a stationary motion but with zero angular momentum. We study the structure of the rate of variation of the atomic energy. The intention is to clarify in a quantitative way the effect of the distinguished contributions of vacuum fluctuations and radiation reaction on spontaneous excitation and on spontaneous emission of atoms. In particular, we are interested in the comprehension of the combined action of the different physical processes underlying the Hawking effect in the scenario of rotating black holes as well as the Unruh-Starobinskii effect. We demonstrate that, in the case of static atoms, spontaneous excitation is also connected with the Unruh-Starobinskii effect, but only in the case of the quantum field prepared in the Frolov-Thorne vacuum state. In addition, we show that, in the ZAMOs perspective, the Boulware vacuum state contains an outward flux of particles as a consequence of the black-hole superradiance. The possible relevance of the findings in the present work is discussed.
\end{abstract}


\maketitle

\section{Introduction}
\label{intro}

The Kerr metric describes the geometry of empty spacetime around a rotating uncharged axially-symmetric black hole~\cite{frolov}. In fact, uniqueness theorems within the general-relativity framework imply that all isolated, vacuum black holes in the Universe are described by the two-parameter Kerr family~\cite{ellis}. An interesting particularity in connection with such rotating black holes is the verification of frame dragging: At close distances from the event horizon, objects must rotate with the black hole. The region where this takes place is called the ergosphere. Any stationary and axisymmetric spacetime with an event horizon must possess an ergoregion~\cite{cardoso}. In turn, the ergoregion can exist in rotating spacetimes with no horizons, such as rapidly rotating neutron stars~\cite{rot1, rot2}. 

An important concept associated with rotating black holes is the superradiance, a classical circumstance of wave amplification~\cite{zeldovich,misner,teu3}. Indeed, we have witnessed a vigorous revival of studies of black-hole superradiance in different branches of physics. One example, of immense current interest, is the investigation of the usage of superradiant instabilities to constrain the mass of ultralight degrees of freedom~\cite{dub,pani,brito}, with substantial implications to dark-matter searches and to physics beyond the Standard Model. It is known that superradiant instabilities of bosons have significant phenomenological effects, usually associated with certain electromagnetic and gravitational waves emission from astrophysical black holes. In turn, superradiance is found to be pivotal in resolving the stability of black holes. For more discussions on the state of the art of black-hole superradiance and its recent applications see Ref.~\cite{review} and references cited therein.

On the astrophysical side, one of the main topics of important ongoing research concerns the study of relativistic jets emitted by rotating black holes~\cite{meier}. The driving machinery that supports these events is still under dispute to some extent. We remark that one of the most successful experimental methods to analyze the engines generating the jets is to specify the jet composition at radii in which they can be seen directly~\cite{mirabel,blandford}. In turn, the most energetic jets are perceived in the so-called active galactic nuclei, and they are supposed to be the outcome of accretion of matter by supermassive rotating black holes~\cite{Lynden}. The energy which is necessary for the acceleration of these streams of matter is believed to come from gravitational binding energy or from the black hole's rotational energy, or from even both. With respect to the latter, there are two well known formalisms aiming at understanding how energy is transferred from a rotating black hole to a jet, namely: the Blandford-Znajek process (BZ)~\cite{blandford2}, which considers the black hole immersed in magnetic fields, and the Penrose mechanism~\cite{penrose}.  Currently the BZ process is widely considered as one of the most auspicious mechanisms for powering relativistic jets from black holes. 

It has been shown that all such processes (superradiance, BZ process and the Penrose mechanism) could be considered to be akin to each other since they rest on the same physical foundation: extraction of the black hole's rotational energy. In other words, the necessary and sufficient condition for all these processes to take place is the absorption of negative energy and angular momentum by the black hole~\cite{kom,lasota}. For instance, for the BZ process when $0 < \Omega_{\textrm{EM}} < \Omega_{H}$, where $\Omega_{H}$ is the angular velocity of the black hole and $\Omega_{\textrm{EM}}$ is the angular velocity of the electromagnetic field, one verifies a net radial negative energy and angular momentum at the horizon~\cite{review}. This means that energy and angular momentum are being extracted from the black hole and, in an astrophysical context, might contribute to the energy released in the relativistic jets. The above condition is very similar to the usual condition for superradiance to occur: 
$\omega < m\Omega_{H}$, where $\omega$ is the frequency of the incident radiation and $m$ is the azimuthal number with respect to the rotation axis. On the other hand, we stress the importance of the existence of the event horizon. Indeed, as discussed in Ref.~\cite{review}, if the spacetime is described by a rotating black-hole geometry, both superradiance and Penrose's process can take place. In spite of that, a perfect-fluid star may allow for Penrose's process yet not for superradiance~\cite{penner}. 

The Unruh-Starobinskii process is the quantum analog of the superradiance~\cite{sta,unruh-sta,davies}. It predicts that particles should be produced by the rotational motion of the black hole. This specific vacuum instability predates the discovery of the Hawking effect which describes how black holes cause thermal particle production~\cite{haw1}. Even though such processes have some connection (the existence of an event horizon), one must not refrain from observing the important differences between them. Notwithstanding, it is of significant importance to understand the emergence of such phenomena (and the possible relationship between them) by studying the experiences of a model particle detector interacting with a radiation field in the exterior region of a Kerr black hole. Based on the arguments presented above, investigations involving the Unruh-Starobinskii effect may be relevant to many different areas of research, including the studies of astrophysical processes which are important in the description of the emission of jets from accretion disks around supermassive rotating black holes.

The aim of the present paper is to contribute to the foregoing discussion under an alternative perspective. We propose to investigate radiative processes of atoms in Kerr spacetime. Concerning spontaneous emission of atoms in Minkowski spacetime, quantum field theoretical appraisals reveal the importance of taking into account the vacuum fluctuations and radiation reaction in such processes. In fact, the contribution of these effects are found to be depended on the ordering chosen for commuting atomic and field operators~\cite{mil1}. As carefully demonstrated by Dalibard, Dupont-Roc, and Cohen-Tannoudji (DDC)~\cite{cohen2,cohen3}, there exists a preferred operator ordering with which one can clearly interpret both effects as having an independent physical meaning. We remark that such a formalism (henceforth called DDC formalism) enables one to understand the Unruh effect as the interplay between these two physical events~\cite{aud1,aud2}. Indeed, in such references it has been shown that for an uniformly accelerated atom initially prepared in the ground state interacting with a quantum field in the Minkowski vacuum the balance between vacuum fluctuations and radiation reaction is no longer complete, and transitions to excited states become realizable. This ties the existence of Unruh effect to the spontaneous excitation of atoms in the Minkowski vacuum. For a Schwarzschild black hole the Hawking effect can also be interpreted in the same way~\cite{china5,china6,china}. Other interesting physical applications of the DDC formalism can be found in Refs.~\cite{china1,china2,rizz}. On the other hand, in recent investigations regarding quantum entanglement, the DDC formalism has proved to be a central approach in order to comprehend the radiative processes involving entangled atoms in Minkowski spacetime and also in a black-hole background, as demonstrated in Refs.~\cite{ng1,ng2,ng3}. For related work see the recent Refs.~\cite{ref1,ref2} in which the authors investigate the resonance interaction between entangled atoms in accelerated motion.

As remarked above, in the situation of a rotating black-hole spacetime there emerges two important physical processes, the Hawking and the Unruh-Starobinskii effects. Our motivation here is to seek an interpretation of both effects as the result of the cooperation between vacuum fluctuations and radiation reaction, in the light of the DDC formalism. On the one hand, by providing an alternative deduction of the Hawking radiation for rotating black holes we are considering a generalization (in a certain sense) of the studies of Refs.~\cite{china5,china6,china} which have revealed the relationship between the Hawking radiation and the spontaneous excitation of atoms outside a static, spherically symmetric black hole. On the other hand, since, as argued above, the Unruh-Starobinskii effect  is the quantum counterpart of superradiance, by interpreting the former as the interplay between vacuum fluctuations and radiation reaction one may be led to conjecture that a similar physical interpretation is available to the quantum analogue of all astrophysical processes related with superradiance, such as the BZ process. In other words, we give a possible realization of the role played by different physical processes underlying the quantum description of mechanisms that could be behind the machinery powering relativistic jets from black holes. In this way we believe that the present work furnishes indications on the relevant connection between three current research topics of major importance, namely relativistic quantum information theory, black-hole superradiance and relativistic jets associated with rotating black holes.

When one investigates radiative processes of atoms coupled with vacuum fluctuations of a quantum fields in a curved spacetime, a delicate issue emerges which is related with the choice for the vacuum state of the quantum fields. This has a long story in the researches in semiclassical gravity. We do not propose to give a thorough discussion on the subject since this has been done in many different places, see for instance the Ref.~\cite{birrel}. Let us consider the case of a Schwarzschild spacetime, whose metric is the vacuum solution to the Einstein's equations that represents the gravitational field outside a spherically symmetric massive body. In summay, the calculation of expectation values of physical observables, such as the energy-momentum tensor of matter, has led us to the following physical interpretation. If one is willing to study atomic radiative processes in the exterior vicinity of very compact objects, then the appropriate choice of vacuum state for quantum fields is the Boulware vacuum. On the other hand, in the studies of gravitational collapse of massive rotating bodies the relevant vacuum state is the Unruh vacuum state. In turn, by considering the situation in which there is a steady-state thermal equilibrium between the black hole and its surroundings, one is led to work with Hartle-Hawking vacuum state. This does not correspond to our usual notion of a vacuum state. In the case of a rotating black hole, some peculiarities arise that will be briefly discussed in due course. This is related with the existence of superradiant modes. In this case, one may define two kinds of ``Boulware" vacuum states and two associated ``Unruh" vacuum states. However, a true Hartle-Hawking vacuum state for the Kerr black hole cannot be defined. So the best one can do is to conceive the construction of a thermal state that contains most but not all of the properties of a true Hartle-Hawking state. In the literature we have two proposals for such a state, namely the Candelas-Chrzanowski-Howard vacuum state and Frolov-Thorne vacuum state. One might also envisage the Boulware vacuum state as the ``pure vacuum polarization part" of the Hartle-Hawking-like vacuum states mentioned above and hence acquire further insight of the inexorability of Hawking radiation~\cite{sciama}. In this work we also intend to improve our understanding on the physical meaning of each of those proposed vacuum states. 

In this paper we study radiative processes of atoms in a Kerr background. We use the DDC approach. The organization of the paper is as follows. In Section~\ref{model} we discuss the identification of vacuum fluctuations and radiation reaction effect in the rate of variation of atomic observables. We consider for simplicity a two-level atom interacting with a quantum massless scalar field in Kerr spacetime. This simplified assumption will prove to be not much restrictive since it is able to capture all the main features of the case in study. In addition, within the terminology of quantum information theory, such an atomic system describes a qubit. In turn, our two-level atom can also be envisaged as a simplified version of the Unruh-DeWitt particle detector~\cite{unruh,dw2}. In Section~\ref{rate-static} we calculate the rates of variation of atomic energy for an atom placed at fixed radial distances outside a Kerr black hole. We take the quantum field as being prepared in each of the physical vacuum states of interest, namely the Boulware vacuum state, the Unruh vacuum state and the two vacuum states which would be ``equivalent" to the Hartle-Hawking vacuum, the Candelas-Chrzanowski-Howard vacuum state and Frolov-Thorne vacuum state. In Section~\ref{rate-stationary} we extend the results of the previous Section to the situation in which the atom has a stationary motion with a zero angular momentum. Conclusions and final remarks are given in Section~\ref{conclude}. The appendices contain a brief digression on different reference frames of observers in Kerr spacetime as well as lengthy derivations of the correlation functions of the scalar field. In this paper we use units such that $\hbar = c = G = k_B = 1$. 

\section{The coupling of atoms with massless scalar fields in a rotating black-hole spacetime}
\label{model}

Here we are interested in studying the contributions of vacuum fluctuations and radiation reaction for the radiative processes of an atom interacting with a quantum scalar field outside a Kerr black hole. As discussed above, the Kerr metric is the most general stationary asymptotically flat vacuum solution of the Einstein's equations describing a rotating black hole. In the standard Boyer-Lindquist coordinates the Kerr metric is given by~\cite{frolov} 
\bea
ds^2 &=& -\left(1 - \frac{2 M r}{\rho^2}\right)\,dt^2 - \frac{4 M a r \sin^2\theta}{\rho^2}\,dt\,d\phi 
\nn\\
&+&\, \frac{\rho^2}{\Delta}\,dr^2 + \rho^2\,d\theta^2 + \frac{\Sigma}{\rho^2}\,\sin^2\theta\,d\phi^2
\nn\\
&=& -\frac{\rho^2\Delta}{\Sigma}\,dt^2 + \frac{\Sigma}{\rho^2}\,\sin^2\theta\,(d\phi - w dt)^2
\nn\\
&+&\, \frac{\rho^2}{\Delta}\,dr^2 + \rho^2\,d\theta^2
\eea
where $w = - g_{0\phi}/g_{00}$ and:
\bea
\rho^2 &=& r^2 + a^2\cos^2\theta
\nn\\
\Delta &=& r^2 - 2 M r + a^2
\nn\\
\Sigma &=& (r^2 + a^2)^2 - a^2\Delta\sin^2\theta.
\eea
 We are using the convention in which the Minkowski metric is given by: $\eta_{\alpha\beta} = 1, \alpha=\beta=1,2,3$, $\eta_{\alpha\beta} = - 1, \alpha=\beta=0$ and $\eta_{\alpha\beta} = 0,\alpha \neq \beta$.
The components of the inverse metric are 
\bea
g^{00} &=& - \frac{\Sigma}{\rho^2\Delta}
\nn\\
g^{0\phi} &=& - \frac{2 M a r}{\rho^2\Delta}
\nn\\
g^{\phi\phi} &=& \frac{\Delta - a^2\sin^2\theta}{\rho^2\Delta\sin^2\theta}
\nn\\
g^{rr} &=&  \frac{\Delta}{\rho^2}
\nn\\
g^{\theta\theta} &=&  \frac{1}{\rho^2}.
\eea
The Kerr metric is stationary and axially symmetric. It has two commuting Killing vectors which we call $\xi_t$ and $\xi_{\phi}$. The Killing vector $\xi_t$ generates translation in time whereas $\xi_{\phi}$ is a generator of rotations. Any linear combination of the Killing vectors with constant coefficients is also a Killing vector. This implies that if one is willing to specify the vectors $\xi_t$ and $\xi_{\phi}$ uniquely one finds the necessity of imposing further conditions: $\xi_{t}$ is the Killing vector which is timelike at infinity with norm $\xi_{t}^2 = -1$; and the integral lines of the Killing vector field $\xi_{\phi}$ are closed. The Kerr metric has two parameters, the mass $M$ and rotation parameter $a \leq M$, connected with the angular momentum of the black hole $J$ by $a = J/M$. 

The Killing vector $\xi_t$ becomes spacelike at all points close to the boundary of the rotating black hole. This takes place in the region where $\xi_t^2 = g_{00} > 0$. The set of all points of this region which is outside the rotating black hole is called ergosphere. In fact, $\xi_t$ becomes null at the boundary of the ergosphere. This is the static limit surface in which the component $g_{00}$ vanishes. One has that:
\beq
r_{st} = M + \sqrt{M^2 - a^2\cos^2\theta}.
\eeq
On the other hand, the equation $\Delta = 0$ has two roots:
\beq
r_{\pm} = M \pm \sqrt{M^2 - a^2}.
\eeq
These are just coordinate singularities, but the Kerr spacetime is truly singular at $\rho^2 = 0$. The surface determined by $r_{+}$ defines the event horizon. The angular velocity of the event horizon is given by
\beq
\Omega_{H} = \frac{a}{r_{+}^{2} + a^2}.
\label{angular-bh}
\eeq

Let us consider an atom, described hypothetically by a two-level system, following a stationary trajectory in the Kerr spacetime with the geometry described above. The stationary trajectory condition guarantees the existence of stationary states. The Hamiltonian of the two-level atom can then be written as
\beq
H_A(\tau) = \frac{\omega_0}{2}\,\sigma^z(\tau),
\label{ha-kerr}
\eeq
where $\tau$ stands for the proper time of the two-level atom and $\sigma^z = | e \rangle\langle e | - | g \rangle\langle g |$. Here $|g\rangle$ and $|e\rangle$ denote the ground and excited states of an isolated qubit, with energies $-\,\omega_0/2$ and $+\,\omega_0/2$, respectively. Here we consider that the atomic system is coupled with a quantum massless scalar field. The Hamiltonian $H_F$ of the free scalar field can be obtained in the usual way. One has that
\beq
H_F(\tau) = \int\,d{{\bf k}}\, \omega_{{\bf k}}\,a^{\dagger}_{{\bf k}}(t(\tau))a_{{\bf k}}(t(\tau))\,\frac{dt}{d\tau},
\label{hf-kerr}
\eeq
where $a^{\dagger}_{{\bf k},\lambda}, a_{{\bf k},\lambda}$ are the usual creation and annihilation operators of the quantum field and we have neglected the zero-point energy. In addition, ${\bf k}$ labels the wave vector of the field modes. Furthermore, we also assume that the presence of a gravitational field does not affect substantially the physical consequences in considering the interaction between the atom and the field. Hence one has that the Hamiltonian which describes the interaction between the two-level atom and the field is given by
\beq
H_{I}(\tau) =  \lambda\,m(\tau) \varphi(x(\tau)),
\label{hi-kerr}
\eeq
where $\lambda$ is a coupling constant assumed to be small and $m(\tau)$ is the monopole moment operator for the qubit. This is given by
\beq
m(\tau) = \sigma^{+}(\tau) + \sigma^{-}(\tau) = | e\rangle\langle g|\,e^{i\omega_0\tau} 
+ |g \rangle\langle e|\,e^{-i\omega_0\tau}.
\label{mon}
\eeq
The associated Heisenberg equations of motion with respect to $\tau$ can be derived from the total Hamiltonian given by $H(\tau) = H_A(\tau) + H_F(\tau) + H_I(\tau)$. In this situation one usually separates the solutions of the equation of motion in two parts, namely: The free part (independent of the details of the interaction between atoms and fields) and the source part (which describes the interaction between atoms and fields). That is, for atomic operators one finds (the superscripts $f,s$ stand for the free and source part, respectively): $\sigma^z(\tau) = \sigma^{z,f}(\tau) + \sigma^{z,s}(\tau)$ and also $m(\tau) = m^{f}(\tau) + m^{s}(\tau)$, whereas for field operators one has that $a_{{\bf k}}(t(\tau)) = a^{f}_{{\bf k}}(t(\tau)) + a^{s}_{{\bf k}}(t(\tau))$. In consequence, one can also write $\varphi(t(\tau)) = \varphi^{f}(t(\tau)) + \varphi^{s}(t(\tau))$. Nonetheless, this calculation produces an ambiguity concerning products of atomic and field operators which implies that one must choose an operator ordering when considering the action of $\varphi^{f}$ (responsible for the effects of vacuum fluctuations) and $\varphi^s$ (which originates the radiation-reaction contribution) individually. This has led to a particular prescription which enables to interpret the effects of such phenomena as independent physical processes~\cite{cohen2,cohen3}. This is primarily the DDC formalism mentioned above. The idea is to evaluate $d {\cal A}/ dt$, where ${\cal A}$ is any given atomic observable, identify the part which is due to the interaction with the field, uncover the contributions of vacuum fluctuations and radiation reaction to this part and then take the expectation value of the resulting quantities. The latter consists of two procedures: first one considers vacuum expectation values regarding quantum field operators; afterwards one takes the expectation value of the associated expressions in a state $|\nu\rangle$, with energy $\nu$. In the present case such a state is either $|g\rangle$ or $|e\rangle$. 

Let us present the contributions of vacuum fluctuations and radiation reaction in the evolution of the atom's energies. Hence we take ${\cal A} = H_A$ which is given by Eq.~(\ref{ha-kerr}). By employing an usual perturbative expansion and taking into account only terms up to order $\lambda^2$, the vacuum-fluctuation contribution reads
\beq
\Biggl\langle \frac{d H_A}{dt} \Biggr\rangle_{VF} = \frac{i\lambda^2}{2}\int_{\tau_0}^{\tau}d\tau'
D(x(\tau),x(\tau'))\frac{\partial}{\partial\tau}\chi(\tau,\tau'),
\label{vfha3}
\eeq
where the notation $\langle (\cdots) \rangle = \langle 0,\nu |(\cdots)| 0, \nu \rangle$ has been employed ($| 0\rangle$ is the vacuum state of the field). In the above:
\beq
\chi(\tau,\tau') = \langle \nu | [m^{f}(\tau),m^{f}(\tau')]| \nu \rangle,
\label{susa}
\eeq
is the linear susceptibility of the atom in the state $|\nu\rangle$ and 
\beq
D(x(\tau),x(\tau')) = \langle 0 |\{\varphi^{f}(x(\tau)),\varphi^{f}(x(\tau'))\}| 0 \rangle,
\eeq
$a, b =1,2$ is the Hadamard's elementary function. On the other hand, for the radiation-reaction contribution, one has:
\beq
\Biggl\langle \frac{d H_A}{dt} \Biggr\rangle_{RR} = \frac{i\lambda^2}{2}\int_{\tau_0}^{\tau}d\tau'
\Delta(x(\tau),x(\tau'))\frac{\partial}{\partial\tau} C(\tau,\tau'),
\label{rrha3}
\eeq
where
\beq
C(\tau,\tau') = \langle \nu | \{m^{f}(\tau),m^{f}(\tau')\}| \nu \rangle,
\label{cora}
\eeq
is the symmetric correlation function of the atom in the state $|\nu\rangle$ and 
\beq
\Delta(x(\tau),x(\tau')) = \langle 0 |[\varphi^{f}(x(\tau)),\varphi^{f}(x(\tau'))]| 0 \rangle,
\eeq
is the Pauli-Jordan function. Likewise, observe that such a formalism enables one to discuss the interplay between vacuum fluctuations and radiation reaction in the radiative processes of atoms. As emphasized in many works, $\chi$ and $C$ characterize only the qubit itself. The explicit forms of such quantities are given by
\bea
\chi(\tau,\tau') &=& \sum_{\nu'}\,{\cal M}(\nu\to\nu')
\nn\\
&\times&\,\biggl[e^{i\Delta\nu(\tau - \tau')} - e^{-i\Delta\nu(\tau - \tau')}\biggr],
\label{susa1}
\eea
and
\bea
C(\tau,\tau') &=& \sum_{\nu'}\,{\cal M}(\nu\to\nu')
\nn\\
&\times&\,\,\biggl[e^{i\Delta\nu(\tau - \tau')} + e^{-i\Delta\nu(\tau - \tau')}\biggr],
\label{cora1}
\eea
where $\Delta \nu = \nu - \nu'$ and the sum extends over the complete set of atomic states. In addition, we have defined the quantity ${\cal M}(\nu\to\nu') = |\langle \nu |m^{f}(0)| \nu' \rangle|^2$. In the situation considered in this paper, 
$\Delta \nu = \omega_0$ for the transition $|e\rangle \to |g\rangle$, while the transition $|g\rangle \to |e\rangle$ has $\Delta \nu = -\omega_0$. In both cases, ${\cal M}(e \to g) = {\cal M}(g \to e) = 1$. We now proceed to calculate the rate of variation of energy for two different classes of world lines for the qubit.

\section{Rate of variation of energy for a static atom}
\label{rate-static}

In the case treated in this Section our two-level atom follows the world line given by $x^{\mu}(\tau) = (t(\tau),r,\theta,\phi)$, where $r, \theta, \phi$ are constants and $t(\tau) = \int d\tau u^0 = \int d\tau |g_{00}|^{-1/2} = \tau |g_{00}|^{-1/2}$, and one has that $g_{00}(r,\theta) = -(1 - 2 M r/\rho^2)$. In this situation we consider the atom to remain in the region $r > r_{st}$. For more details on static observers in Kerr spacetime we refere the reader the Appendix~\ref{A} which contains brief comments on the  different families of observers of interest in Kerr spacetime. We also consider the scalar field prepared in each of the physical vacuum states of interest, namely the Boulware vacuum state, the Unruh vacuum state and the two candidates for the vacuum state equivalent to the Hartle-Hawking vacuum, known as the Candelas-Chrzanowski-Howard vacuum state and Frolov-Thorne vacuum state. Interesting discussions on the difficulties of defining a state equivalent to the Hartle-Hawking vacuum can be found in Ref.~\cite{ote}. In addition, in Appendix~\ref{B} we collect all the correlation functions of the scalar field that will be important in what follows.

\subsection{The Boulware vacuum states}

In Schwarzschild spacetime, the Boulware vacuum has a close similarity to the concept of an empty state at large radii, but it
has pathological behavior at the horizon~\cite{boul}. The Boulware vacuum is the appropriate choice of vacuum state for
quantum fields in the vicinity of an isolated, cold neutron star. On the other hand, in Kerr spacetime the discussion is more involved: The existence of superradiant modes render the discussion on positive frequencies more intricate. The definition of physical states with certain properties is straightforward only along a given Cauchy surface~\cite{ote}. Here we consider the two Cauchy surfaces, ${\cal J}^{-} \bigcup {\cal H}^{-}$ and ${\cal J}^{+} \bigcup {\cal H}^{+}$. For the first (latter) we define a past (future) Boulware vacuum state. Such states do not strictly agree with the common idea of a Boulware state in Schwarzschild spacetime as the most empty state at infinity; this is a direct consequence of the Unruh-Starobinskii effect~\cite{unruh-sta}. The non-existence of a Boulware vacuum state in the precise sense as employed in Schwarzschild spacetime is intimately connected with the fact that a true Hartle-Hawking state cannot be defined on Kerr spacetime~\cite{wald}. For more discussions on the Boulware vacuum states, we refer the reader the Appendix~\ref{B}.

Let us calculate the rate of variation of atomic energy in the Boulware vacuum states. Details concerning the derivation of all the relevant correlation functions of the massless scalar field which appears in Eqs.~(\ref{vfha3}) and~(\ref{rrha3}) can be found in the Appendix~\ref{B}. The contributions of vacuum fluctuations are then given by
\bea
\Biggl\langle \frac{d H_A}{dt} \Biggr\rangle_{VF} &=& -\lambda^2\sum_{l,m}\,
\sum_{\nu'}\,{\cal M}(\nu\to\nu')\Delta\nu
\nn\\
&\times&\,\int_{-\Delta\tau}^{\Delta\tau}du\,\Biggl[\int_{0}^{\infty}\,d\omega\,{\cal P}^{+}_{\omega l m}(r, \theta)\, 
\cos(w u) 
\nn\\
&+&\, \int_{0}^{\infty}\,d\bar{\omega} {\cal P}^{-}_{\omega l m}(r, \theta)\,
\cos(w u) \Biggr]\,e^{i\Delta\nu u},
\label{vf-boul}
\eea
where $\Delta\tau = \tau - \tau_0$, $w = w(\omega) = \omega |g_{00}|^{-1/2}$ and
\bea
{\cal P}^{\pm}_{\omega l m}(r, \theta) = \frac{|S_{\omega l m}(\cos\theta)|^2\,|R^{\pm}_{\omega l m} (r)|^2}
{8\pi^2\omega^{\pm}(r^2 + a^2)},
\eea
with $\omega^{+} = \omega$ and $\omega^{-} = \bar{\omega} = \omega - m\Omega_{H}$. We considered only the past modes in order to derive expression~(\ref{vf-boul}) but a similar analysis can be carried out for the future modes. Actually, one can use the asymptotic forms of the modes presented in the Ref.~\cite{ote} in order to discuss the behavior of ${\cal P}^{\pm}$ near the event horizon and also far away from the black hole. One gets:
\begin{widetext}
\begin{eqnarray}
{\cal P}^{+}_{\omega l m}(r, \theta) &\sim& \frac{|S_{\omega l m}(\cos\theta)|^2}
{8\pi^2\omega}\,
	\begin{cases}
               0 \hspace{90pt} \textrm{at ${\cal H}^{-}$}\\
               (r^2 + a^2)^{-1} \hspace{45pt} \textrm{at ${\cal J}^{-}$}\\
               (r_{+}^2 + a^2)^{-1}\,|{\cal T}^{+}_{\omega l m}|^2 \hspace{11pt} \textrm{at ${\cal H}^{+}$}\\
               (r^2 + a^2)^{-1}\,|{\cal R}^{+}_{\omega l m}|^2 \hspace{10pt} \textrm{at ${\cal J}^{+}$} 
	\end{cases}
\label{pasymp1}
\end{eqnarray}
and
\begin{eqnarray}
{\cal P}^{-}_{\omega l m}(r, \theta) &\sim& \frac{|S_{\omega l m}(\cos\theta)|^2}
{8\pi^2(\omega - m\Omega_{H})}\,
	\begin{cases}
               (r_{+}^2 + a^2)^{-1} \hspace{42pt} \textrm{at ${\cal H}^{-}$}\\
               0 \hspace{89pt} \textrm{at ${\cal J}^{-}$}\\
              (r_{+}^2 + a^2)^{-1}\, |{\cal R}^{-}_{\omega l m}|^2 \hspace{7pt} \textrm{at ${\cal H}^{+}$}\\
               (r^2 + a^2)^{-1}\,|{\cal T}^{-}_{\omega l m}|^2 \hspace{12pt} \textrm{at ${\cal J}^{+}$}, 
	\end{cases}
\label{pasymp2}
\end{eqnarray}
\end{widetext}
where ${\cal R}, {\cal T}$ are the reflection and transmission coefficients, respectively. In principle, one could derive more detailed expressions for such asymptotic expressions but these simple WKB forms are suited enough to the present purposes.

Coming back to our problem, the time integrals appearing in Eq.~(\ref{vf-boul}) can be solved in a straightforward way and one finds
\bea
\Biggl\langle \frac{d H_A}{dt} \Biggr\rangle_{VF} &=& -2\lambda^2\sum_{l,m}\,
\sum_{\nu'}\,{\cal M}(\nu\to\nu')\Delta\nu
\nn\\
&\times&\,\Biggl\{\int_{0}^{\infty}\,d\omega\,{\cal P}^{+}_{\omega l m}(r, \theta)\, {\cal G}_{1}(w, \Delta\nu, \Delta\tau) 
\nn\\
&+& \int_{0}^{\infty}\,d\bar{\omega} {\cal P}^{-}_{\omega l m}(r, \theta)\,{\cal G}_{1}(w, \Delta\nu, \Delta\tau)\Biggr\}
\label{vf-boul-finite}
\eea
where
\bea
{\cal G}_{1}(w, \Delta\nu, \Delta\tau) &=& \frac{1}{w^2-\Delta\nu^2}
\nn\\
&\times&\left[w\sin (w\Delta\tau) \cos (\Delta\nu\Delta \tau) \right.
\nn\\
&-&\, \left. \Delta\nu  \cos(w\Delta\tau) 
\sin (\Delta\nu\Delta\tau)\right].
\eea
For sufficiently large $\Delta \tau$, the integrals over $\omega$ can be explicitly solved with the help of simple Fourier transforms. Thence, with $\Delta\nu\,(u^{0})^{-1} = \widetilde{\Delta\nu}$ (here in the static case, one has that $u^{0} = |g_{00}|^{-1/2}$) one gets:
\bea
\Biggl\langle \frac{d H_A}{dt} \Biggr\rangle_{VF} &=& -\pi\lambda^2\sum_{l}\,
\sum_{\nu'}\,{\cal M}(\nu\to\nu')\widetilde{\Delta\nu}
\nn\\
&\times&\,\theta(\widetilde{\Delta\nu})\,{\cal K}_{l}(\widetilde{\Delta\nu},r, \theta)
\nn\\
&+&\pi\lambda^2\sum_{l}\,
\sum_{\nu'}\,{\cal M}(\nu\to\nu')|\widetilde{\Delta\nu}|
\nn\\
&\times&\,\theta(-\widetilde{\Delta\nu})\,{\cal K}_{l}(|\widetilde{\Delta\nu}|,r, \theta),
\label{vf-boul2}
\eea
where we have defined:
\bea
{\cal K}_{l}(\widetilde{\Delta\nu},r, \theta) &=& \sum_{m} \left({\cal P}^{+}_{\widetilde{\Delta\nu} l m}(r, \theta) + {\cal P}^{-}_{\widetilde{\Delta\nu} l m}(r, \theta)\right)
\nn\\
&\times&\,\theta(\widetilde{\Delta\nu} - m\Omega_{H})
\nn\\
&+&\,\sum_{m>0}\,{\cal P}^{+}_{\widetilde{\Delta\nu} l m}(r, \theta)\,\theta(-\widetilde{\Delta\nu} + m\Omega_{H}).
\eea
It becomes clear to note that vacuum fluctuations tend to excite ($\widetilde{\Delta \nu}_i < 0 \Rightarrow \langle d H_A/dt \rangle_{VF} > 0$) as well as deexcite ($\widetilde{\Delta \nu}_i > 0 \Rightarrow \langle d H_A/dt \rangle_{VF} < 0$) the atomic system.

Now let us present the contribution from radiation reaction. It reads
\bea
\Biggl\langle \frac{d H_A}{dt} \Biggr\rangle_{RR} &=& i\lambda^2\sum_{l,m}\,\sum_{\nu'}\,
{\cal M}(\nu\to\nu')\Delta\nu
\nn\\
&\times&\,\int_{-\Delta\tau}^{\Delta\tau}du\,
\Biggl[\int_{0}^{\infty}\,d\omega\,{\cal P}^{+}_{\omega l m}(r, \theta)\,\sin(w u) 
\nn\\
&+&\, \int_{0}^{\infty}\,d\bar{\omega} {\cal P}^{-}_{\omega l m}(r, \theta)\,\sin(w u) \Biggr]\,e^{i\Delta\nu u}.
\label{rr-boul}
\eea
As above, such an expression is equally valid for both past and future modes. After solving the time integrals one gets:
\bea
\Biggl\langle \frac{d H_A}{dt} \Biggr\rangle_{RR} &=& -2\lambda^2\sum_{l,m}\,\sum_{\nu'}\,
{\cal M}(\nu\to\nu')\Delta\nu
\nn\\
&\times&\,\Biggl\{\int_{0}^{\infty}\,d\omega\,{\cal P}^{+}_{\omega l m}(r, \theta)\,
{\cal G}_{2}(w, \Delta\nu, \Delta\tau) 
\nn\\
&+&\int_{0}^{\infty}\,d\bar{\omega} {\cal P}^{-}_{\omega l m}(r, \theta)\,
{\cal G}_{2}(w, \Delta\nu, \Delta\tau)\Biggr\}
\label{rr-boul-finite}
\eea
where\\
\bea
{\cal G}_{2}(w, \Delta\nu, \Delta\tau) &=& \frac{1}{w^2 - \Delta\nu^2}
\nn\\
&\times&\,\left[\Delta\nu\sin(w\Delta\tau) \cos(\Delta\nu\Delta\tau) \right.
\nn\\
&-&\, \left. w \cos(w\Delta\tau) \sin(\Delta\nu\Delta\tau)\right].
\eea
Again for sufficiently large $\Delta \tau$, the integrals over $\omega$ can be explicitly solved and the result is
\bea
\Biggl\langle \frac{d H_A}{dt} \Biggr\rangle_{RR} &=& -\pi\lambda^2\sum_{l}\,
\sum_{\nu'}\,{\cal M}(\nu\to\nu')\widetilde{\Delta\nu}
\nn\\
&\times&\,\theta(\widetilde{\Delta\nu})\,{\cal K}_{l}(\widetilde{\Delta\nu},r, \theta)
\nn\\
&-&\,\pi\lambda^2\sum_{l}\,
\sum_{\nu'}\,{\cal M}(\nu\to\nu')|\widetilde{\Delta\nu}|
\nn\\
&\times&\,\theta(-\widetilde{\Delta\nu})\,{\cal K}_{l}(|\widetilde{\Delta\nu}|,r, \theta).
\label{rr-boul2}
\eea
Observe that the effect of radiation reaction is to induce a lowering in the atomic energy $\langle d H_A/dt \rangle_{VF} < 0$ independent of how the qubit was initially prepared.

For completeness, let us present the total rate for a finite $\Delta\tau$. It is given by the sum of the vacuum-fluctuation and radiation-reaction contributions, Eqs.~(\ref{vf-boul-finite}) and~(\ref{rr-boul-finite}), respectively. One has, explicitly:
\bea
\Biggl\langle \frac{d H_A}{dt} \Biggr\rangle_{tot} &=& -2\lambda^2\sum_{l,m}\,
\sum_{\nu'}\,{\cal M}(\nu\to\nu')\Delta\nu
\nn\\
&\times&\,\Biggl\{\int_{0}^{\infty}\,d\omega\,{\cal P}^{+}_{\omega l m}(r, \theta)\, 
\nn\\
&\times&\left[{\cal G}_{1}(w, \Delta\nu, \Delta\tau) + {\cal G}_{2}(w, \Delta\nu, \Delta\tau)\right] 
\nn\\
&+& \int_{0}^{\infty}\,d\bar{\omega} {\cal P}^{-}_{\omega l m}(r, \theta)\,
\nn\\
&\times& \left[{\cal G}_{1}(w, \Delta\nu, \Delta\tau) + {\cal G}_{2}(w, \Delta\nu, \Delta\tau)\right]\Biggr\}.
\label{tot-boul-finite}
\eea
The result clearly shows that it is possible to excite the static atom in the ground state for a finite observation time. However, the excited state lasts only a finite duration, and at late observation times $\Delta \tau$ a transition to the excited state is forbidden. Indeed, for $\Delta\tau \to \infty$, $\widetilde{\Delta\nu} > 0$ and $\widetilde{\Delta\nu} > m\Omega_{H}$
\bea
\Biggl\langle \frac{d H_A}{dt} \Biggr\rangle_{tot} &=& -2\pi\lambda^2\,\widetilde{\omega}_0
\nn\\
&\times&\,\sum_{l,m}\biggl[{\cal P}^{+}_{\widetilde{\omega}_0 l m}(r, \theta) 
+ {\cal P}^{-}_{\widetilde{\omega}_0 l m}(r, \theta)\biggr]
\eea
and, for $\widetilde{\Delta\nu} > 0$ and $\widetilde{\Delta\nu} < m\Omega_{H}$ (positive $m$), with $\Delta\tau \to \infty$:
\beq
\Biggl\langle \frac{d H_A}{dt} \Biggr\rangle_{tot} = -2\pi\lambda^2\,\widetilde{\omega}_0\sum_{l,m>0}
\,{\cal P}^{+}_{\widetilde{\omega}_0 l m}(r, \theta),
\eeq
where $\omega_0\,(u^{0})^{-1} = \widetilde{\omega}_0$. On the other hand, the total rate is zero for $\widetilde{\Delta\nu} < 0$ and $\Delta\tau \to \infty$. The interpretation of such results are straightforward: for an atom prepared in the ground state there is a fine tuning between vacuum fluctuations and radiation reaction which prevents the ocurrence of spontaneous excitation to higher levels. In turn, for an atom in the excited state, vacuum fluctuations and radiation reaction concur with equal portions to the spontaneous emission. This physical interpretation is analogous to that usually given for a static atom interacting with a quantum field in the Minkowski vacuum. Hence, even though one can assert that the past Boulware vacuum state does not agree with the concept of a state which is most empty at infinity, in contrast with the Schwarzschild spacetime, we see that static, ground-state atoms coupled with quantum fields in Boulware vacuum outside a Kerr black hole are stable. This point will be better understood when we duly consider the case of a zero angular momentum atom. On the other hand, note that the spontaneous emission rate in the past Boulware vacuum is clearly different from that of an inertial atom in the Minkowski vacuum due to the presence of the quantities ${\cal P}^{\pm}_{\widetilde{\omega}_0 l m}$. 

In order to gain further insight, let us study the behavior of the rate for $r \to \infty$, which is obtained by using the asymptotic form for the functions ${\cal P}^{\pm}$ given by Eqs.~(\ref{pasymp1}),~(\ref{pasymp2}). In particular, for $\widetilde{\omega}_0 \gtrsim m\Omega_{H}$, the decay rate may acquire arbitrarily high absolute values at ${\cal J}^{+}$ due to the contribution coming from ${\cal P}^{-}$. Also, for $\widetilde{\omega}_0 < m\Omega_{H}$ (positive $m$), $|{\cal R}^{+}_{\omega l m}|^2 >1$ and the decay rate is larger at ${\cal J}^{+}$ in comparison with ${\cal J}^{-}$ (in absolute values; also, see comments in the Appendix~\ref{B}): This is a consequence of the superradiance.

\subsection{The Unruh vacuum state}

In Schwarzschild spacetime, the Unruh vacuum state is the adequate choice of vacuum state which is most relevant to the gravitational collapse of a massive spherically symmetric body. At spatial infinity this vacuum is equivalent to an outgoing flux of black-body radiation at the black-hole temperature~\cite{sciama}. In turn, in Kerr spacetime one has two ``types" of Unruh vacuum states: one defines a past (future) Unruh state as that state empty at ${\cal J}^{-}$ (${\cal J}^{+}$) but with modes on ${\cal H}^{-}$ (${\cal H}^{+}$) thermally populated. Since it is the past Unruh state that mimics the state showing up at late times from the collapse of a star to form a black hole, this is the one we consider in this work. For more discussions on the Unruh vacuum state for the Kerr metric, we refer the reader the Appendix~\ref{B}.

Let us proceed to calculate the rate of variation of atomic energy in the (past) Unruh vacuum state. Again all details on the derivation of all the relevant correlation functions of the massless scalar field which appears in Eqs.~(\ref{vfha3}) and~(\ref{rrha3}) are collected in the Appendix~\ref{B}. As above, we first consider the contributions of vacuum fluctuations. These are given by
\bea
\Biggl\langle \frac{d H_A}{dt} \Biggr\rangle_{VF} &=& -\lambda^2\sum_{l,m}\,
\sum_{\nu'}\,{\cal M}(\nu\to\nu')\Delta\nu
\nn\\
&\times&\,\int_{-\Delta\tau}^{\Delta\tau}du\,\Biggl[\int_{0}^{\infty}\,d\omega\,{\cal P}^{+}_{\omega l m}(r, \theta)\, 
\cos(w u) 
\nn\\
&+&\, \int_{0}^{\infty}\,d\bar{\omega} \coth\left( \frac{\pi\bar{\omega}}{\kappa_{+}} \right)
\nn\\
&\times&\,{\cal P}^{-}_{\omega l m}(r, \theta)\,\cos(w u) \Biggr]\,e^{i\Delta\nu u},
\eea
where $\kappa_{+}$ is the surface gravity on the outer horizon, given by Eq.~(\ref{surface-gravity}). One can easily solve the time integrals to obtain
\bea
\Biggl\langle \frac{d H_A}{dt} \Biggr\rangle_{VF} &=& -2\lambda^2\sum_{l,m}\,
\sum_{\nu'}\,{\cal M}(\nu\to\nu')\Delta\nu
\nn\\
&\times&\,\Biggl\{\int_{0}^{\infty}\,d\omega\,{\cal P}^{+}_{\omega l m}(r, \theta)\, 
{\cal G}_{1}(w, \Delta\nu, \Delta\tau) 
\nn\\
&+&\, \int_{0}^{\infty}\,d\bar{\omega} \coth\left( \frac{\pi\bar{\omega}}{\kappa_{+}} \right)
\nn\\
&\times&\,{\cal P}^{-}_{\omega l m}(r, \theta)\,{\cal G}_{1}(w, \Delta\nu, \Delta\tau) \Biggr\}.
\eea
For sufficiently large $\Delta \tau$, the integrals over $\omega$ can be explicitly solved and the result is
\bea
\Biggl\langle \frac{d H_A}{dt} \Biggr\rangle_{VF} &=& -\pi\lambda^2\sum_{l}\,
\sum_{\nu'}\,{\cal M}(\nu\to\nu')\widetilde{\Delta\nu}
\nn\\
&\times&\,\theta(\widetilde{\Delta\nu})\,{\cal U}_{l}(2,\widetilde{\Delta\nu},r,\theta)
\nn\\
&+&\pi\lambda^2\sum_{l}\,
\sum_{\nu'}\,{\cal M}(\nu\to\nu')|\widetilde{\Delta\nu}|
\nn\\
&\times&\,\theta(-\widetilde{\Delta\nu})\,{\cal U}_{l}(2,|\widetilde{\Delta\nu}|,r,\theta)
\eea
where we have defined ${\cal U}_{l}(n,\widetilde{\Delta\nu},r,\theta) = {\cal U}^{1}_{l}(n,\widetilde{\Delta\nu},r,\theta) 
+ {\cal U}^{2}_{l}(n,\widetilde{\Delta\nu},r,\theta)$, with
\bea
&&{\cal U}^{1}_{l}(n,\widetilde{\Delta\nu},r,\theta) = \sum_{m}\Biggl\{{\cal P}^{+}_{\widetilde{\Delta\nu} l m}(r, \theta) 
\nn\\
&&+\, {\cal P}^{-}_{\widetilde{\Delta\nu} l m}(r, \theta)\Biggl[1 + \frac{n}{\exp{\left(\frac{2\pi(\widetilde{\Delta\nu} - m\Omega_{H})}{\kappa_{+}}\right)}-1}\Biggr]\Biggr\}
\nn\\
&&\times\,\theta(\widetilde{\Delta\nu} - m\Omega_{H})
\eea
and
\beq
{\cal U}^{2}_{l}(n,\widetilde{\Delta\nu},r,\theta) = \sum_{m>0}\,{\cal P}^{+}_{\widetilde{\Delta\nu} l m}(r, \theta)\,\theta(-\widetilde{\Delta\nu} + m\Omega_{H})
\eeq
We note the appearance of the thermal terms only for $\widetilde{\Delta\nu} > m\Omega_{H}$ 
(or $|\widetilde{\Delta\nu}| > m\Omega_{H}$). As above, vacuum fluctuations tend to excite an accelerated
atom in the ground state and deexcite it in the excited state. Both processes are heightened by the thermal terms in comparison with the Boulware case only for $\widetilde{\Delta\nu} > m\Omega_{H}$ 
(or $|\widetilde{\Delta\nu}| > m\Omega_{H}$). 

Turning to the contribution from radiation reaction, with the help of the results derived in the Appendix~\ref{B}, one finds the same result as Eq.~(\ref{rr-boul}). Hence we obtain the same results as in the Boulware vacuum state, namely Eq.~(\ref{rr-boul2}). Hence one sees that the radiation reaction does not get any Planckian factor. We expect this to be a typical behavior of the scalar field; for instance, for an electromagnetic field this may not be true~\cite{china}.

For completeness, let us present the total rate, which is obtained by the sum of the vacuum-fluctuation and radiation-reaction contributions. We are particularly interested in the case of large observational times $\Delta\tau$. One gets:
\bea
\Biggl\langle \frac{d H_A}{dt} \Biggr\rangle_{tot} &=& -2\pi\lambda^2\sum_{l}\,
\sum_{\nu'}\,{\cal M}(\nu\to\nu')\widetilde{\Delta\nu}
\nn\\
&\times&\,\theta(\widetilde{\Delta\nu})\,{\cal U}_{l}(1,\widetilde{\Delta\nu},r,\theta)
\nn\\
&+&2\pi\lambda^2\sum_{l,m}\,
\sum_{\nu'}\,{\cal M}(\nu\to\nu')|\widetilde{\Delta\nu}|\,\theta(-\widetilde{\Delta\nu})
\nn\\
&\times&\,\frac{\theta(-\widetilde{\Delta\nu} - m\Omega_{H})
\,{\cal P}^{-}_{|\widetilde{\Delta\nu}| l m}(r, \theta)}{\exp{\left(\frac{2\pi(|\widetilde{\Delta\nu}| - m\Omega_{H})}{\kappa_{+}}\right)}-1}.
\eea
The balance between vacuum fluctuations and radiation reaction that ensures the stability of the atom in its ground
state in the Boulware vacuum no longer exists for $|\widetilde{\Delta\nu}| > m\Omega_{H}$. The last term in the above equation
which gives a positive contribution to the total rate makes the transition of the atom from the ground state to an excited state possible, i.e. excitation spontaneously occurs in the Unruh vacuum outside a rotating black hole. This spontaneous excitation
is the consequence of the Hawking effect, which is now interpreted as the interplay between the two underlying
physical effects. Being more specific, the structure of the rate of change of atomic energy suggests that there is thermal radiation from the rotating black hole only for $|\widetilde{\Delta\nu}| > m\Omega_{H}$; with $|\widetilde{\Delta\nu}| < m\Omega_{H}$ the atom in its ground state do not get excited. Hence there must be a compromise between the angular velocity of the black hole and the energy gap between the states so that one verifies the existence of a thermal radiation; in particular, if the energy gap obeys the superradiant condition, the transition to the excited state is not possible. In turn, it is this thermal radiation that renders the spontaneous excitation possible for a nonsuperradiant energy gap. The temperature of the thermal radiation is given by
\beq
T = \frac{\kappa_{+}}{2\pi}\,|g_{00}|^{-1/2},
\label{temp}
\eeq
with $\kappa_{+}/2\pi$ being the usual Hawking temperature of the black hole. As expected, the rotation of the black hole enters into the thermal spectrum as a chemical potential~\cite{birrel}. In conclusion, the spontaneous excitation of atoms with a gap satisfying the condition $|\widetilde{\Delta\nu}| > m\Omega_{H}$ is the Hawking effect. This is the physical content of the (past) Unruh vacuum state:  It describes a rotating black hole with an outgoing flux of thermal radiation emanated from its event horizon at the Hawking temperature given by Eq.~(\ref{temp}) and satisfying the aforementioned condition.

As in the Boulware case, by using the asymptotic form for the functions ${\cal P}^{\pm}$ given by Eqs.~(\ref{pasymp1}),~(\ref{pasymp2}) one gets the behavior of the rate for $r \to \infty$. In particular, since the thermal terms are multiplied by the gray-body factor ${\cal P}^{-}$ which gives vanishingly contributions at spatial infinity, one gets that the thermal flux flowing from the black hole horizon is strongly backscattered by the curved geometry. Nevertheless, consider 
$\widetilde{\Delta\nu} < 0$: with $|\widetilde{\Delta\nu}| \gtrsim m\Omega_{H}$, it may occur that the rate is not negligible at spatial infinity and spontaneous excitation of the two-level atom is bound to occur in a nontrivial way. Finally, observe that in the case of an extremal black hole ($a = M$), we recover the results of the previous Subsection.

\subsection{The Candelas-Chrzanowski-Howard vacuum state}

It is well known that there does not exist a Hadamard state which is regular everywhere in Kerr spacetime~\cite{wald}. In the absence of such a true Hartle-Hawking vacuum, there were endeavors in the literature in order to define a thermal state with several properties pertained to the Hartle-Hawking state. In this Subsection we shall discuss the results associated with the vacuum state introduced by Candelas, Chrzanowski and Howard~\cite{candelas2}, which is formulated by thermalizing the in and up modes with respect to their natural energy (for a definition of such modes, see Ref.~\cite{ote} and also the Appendix~\ref{B}). Such a vacuum state could be described as the past Hartle-Hawking vacuum; nevertheless, such a state does not respect the simultaneous $t-\phi$ reversal invariance of Kerr spacetime. The Appendix~\ref{B} contains more discussions on the Candelas-Chrzanowski-Howard vacuum state.

Let us calculate the rate of variation of atomic energy in the Candelas-Chrzanowski-Howard vacuum state. Details concerning the derivation of all the relevant correlation functions of the massless scalar field which appears in expressions~(\ref{vfha3}) and~(\ref{rrha3}) can be found in the Appendix~\ref{B}. The contributions of vacuum fluctuations are given by\\
\bea
\Biggl\langle \frac{d H_A}{dt} \Biggr\rangle_{VF} &=& -\lambda^2\sum_{l,m}\,
\sum_{\nu'}\,{\cal M}(\nu\to\nu')\Delta\nu\,\int_{-\Delta\tau}^{\Delta\tau}du
\nn\\
&\times&\,\Biggl[\int_{0}^{\infty}\,d\omega\coth\left( \frac{\pi\omega}{\kappa_{+}} \right)
\,{\cal P}^{+}_{\omega l m}(r, \theta)\cos(w u) 
\nn\\
&+&\, \int_{0}^{\infty}\,d\bar{\omega} \coth\left( \frac{\pi\bar{\omega}}{\kappa_{+}} \right)
\nn\\
&\times&\,{\cal P}^{-}_{\omega l m}(r, \theta)\cos(w u) \Biggr]\,e^{i\Delta\nu u}.
\eea
Similarly as the cases discussed previously, the time integrals can be easily solved and one finds that
\bea
\Biggl\langle \frac{d H_A}{dt} \Biggr\rangle_{VF} &=& -2\lambda^2\sum_{l,m}\,
\sum_{\nu'}\,{\cal M}(\nu\to\nu')\Delta\nu
\nn\\
&\times&\,\Biggl\{\int_{0}^{\infty}\,d\omega\,\coth\left( \frac{\pi\omega}{\kappa_{+}} \right)
\nn\\
&\times&\,{\cal P}^{+}_{\omega l m}(r, \theta)\, 
{\cal G}_{1}(w, \Delta\nu, \Delta\tau) 
\nn\\
&+&\, \int_{0}^{\infty}\,d\bar{\omega} \coth\left( \frac{\pi\bar{\omega}}{\kappa_{+}} \right)
\nn\\
&\times&\,{\cal P}^{-}_{\omega l m}(r, \theta)\,
{\cal G}_{1}(w, \Delta\nu, \Delta\tau) \Biggr\}.
\eea
For large $\Delta\tau \to \infty$, the above expression reduces to a simpler result:
\bea
\Biggl\langle \frac{d H_A}{dt} \Biggr\rangle_{VF} &=& -\pi\lambda^2\sum_{l}\,
\sum_{\nu'}\,{\cal M}(\nu\to\nu')\widetilde{\Delta\nu}
\nn\\
&\times&\,\theta(\widetilde{\Delta\nu})\,{\cal C}_{l}(1,2,\widetilde{\Delta\nu},r,\theta)
\nn\\
&+&\pi\lambda^2\sum_{l}\,
\sum_{\nu'}\,{\cal M}(\nu\to\nu')|\widetilde{\Delta\nu}|
\nn\\
&\times&\,\theta(-\widetilde{\Delta\nu})\,{\cal C}_{l}(1,2,|\widetilde{\Delta\nu}|,r,\theta).
\eea
In order to make the above equation more transparent we have introduced the quantity: 
$$
{\cal C}_{l}(k,n,\widetilde{\Delta\nu},r,\theta)~=~{\cal C}^{1}_{l}(k,n,\widetilde{\Delta\nu},r,\theta)~+~
{\cal C}^{2}_{l}(k,n,\widetilde{\Delta\nu},r,\theta),
$$ 
together with the following definitions:
\begin{widetext}
\bea
{\cal C}^{1}_{l}(k,n,\widetilde{\Delta\nu},r,\theta) &=& \sum_{m}\left[\left(k + \frac{n}{\exp{\left(\frac{2\pi\widetilde{\Delta\nu}}{\kappa_{+}}\right)}-1}\right)
{\cal P}^{+}_{\widetilde{\Delta\nu} l m}(r, \theta) \right.
\nn\\
&+&\,\left. \left(k + \frac{n}{\exp{\left(\frac{2\pi(\widetilde{\Delta\nu} - m\Omega_{H})}{\kappa_{+}}\right)}-1}\right)
{\cal P}^{-}_{\widetilde{\Delta\nu} l m}(r, \theta)\right]\,\theta(\widetilde{\Delta\nu} - m\Omega_{H}),
\eea
and
\beq
{\cal C}^{2}_{l}(k,n,\widetilde{\Delta\nu},r,\theta) = \sum_{m>0}\left(k + \frac{n}{\exp{\left(\frac{2\pi\widetilde{\Delta\nu}}{\kappa_{+}}\right)}-1}\right){\cal P}^{+}_{\widetilde{\Delta\nu} l m}(r, \theta)\,\theta(-\widetilde{\Delta\nu} + m\Omega_{H}).
\eeq
\end{widetext}
Observe the appearance of the thermal contributions in all terms of the above equation. In turn, the contribution from radiation reaction has the same form as Eq.~(\ref{rr-boul}). It is easy to see that we obtain the same results as in the Boulware vacuum state. Hence, the total rate, for large observational times $\Delta\tau$, is given by
\bea
\Biggl\langle \frac{d H_A}{dt} \Biggr\rangle_{tot} &=& -2\pi\lambda^2\sum_{l}\,
\sum_{\nu'}\,{\cal M}(\nu\to\nu')\widetilde{\Delta\nu}
\nn\\
&\times&\,\theta(\widetilde{\Delta\nu})\,{\cal C}_{l}(1,1,\widetilde{\Delta\nu},r,\theta)
\nn\\
&+& 2\pi\lambda^2\sum_{l}\,
\sum_{\nu'}\,{\cal M}(\nu\to\nu')|\widetilde{\Delta\nu}|
\nn\\
&\times&\,\theta(-\widetilde{\Delta\nu})\,{\cal C}_{l}(0,1,|\widetilde{\Delta\nu}|,r,\theta).
\eea
The above results enable one to discuss the physical meaning of the Candelas-Chrzanowski-Howard vacuum state. As in the Unruh vacuum case, the thermal terms also appear, and these terms lead to spontaneous excitation of static atoms in the Candelas-Chrzanowski-Howard vacuum state in the exterior region of the rotating black hole. Besides, the Planckian factor in the total rate of change is a revelation of the thermal nature of the Candelas-Chrzanowski-Howard vacuum state. However, in contrast with the Unruh case, thermal radiation is verified for both cases, $|\widetilde{\Delta\nu}| > m\Omega_{H}$ and $|\widetilde{\Delta\nu}| < m\Omega_{H}$. However, for the contributions coming from the terms proportional to ${\cal P}^{+}$ the rotation of the black hole does not enter into the thermal spectrum. We interpret this as consequence of the fact that this vacuum state does not respect the simultaneous $t-\phi$ reversal invariance of Kerr spacetime. 

From Eqs.~(\ref{pasymp1}),~(\ref{pasymp2}) one gets the asymptotic behavior of the total rate at spatial infinity. Note that terms proportional to ${\cal P}^{\pm}$ are vanishingly small at ${\cal J}^{+}$; yet, at ${\cal J}^{-}$ and 
$|\widetilde{\Delta\nu}| < m\Omega_{H}$, the excitation rate is lower in comparison with ${\cal J}^{+}$ due to the superradiance. However, at ${\cal J}^{-}$ and $|\widetilde{\Delta\nu}| > m\Omega_{H}$, one may expect that the mode sums approach asymptotically the values related to the situation of a static two-level atom immersed in a thermal bath at the Hawking temperature in Minkowski spacetime, since the past Boulware vacuum state corresponds to an absence of particles from ${\cal J}^{-}$ (and ${\cal H}^{-}$). In other words, the Candelas-Chrzanowski-Howard vacuum state is not empty at ${\cal J}^{-}$ but instead it corresponds to a thermal distribution with the Hawking temperature and hence describes at ${\cal J}^{-}$ a rotating black hole in equilibrium with an infinite sea of black-body radiation. This motivates the identification of the Candelas-Chrzanowski-Howard vacuum as a possible candidate to a past Hartle-Hawking vacuum state, according to preceding discussions. Finally, the Boulware-vacuum results are retrieved in the case of an extremal black hole.

\subsection{The Frolov-Thorne vacuum state}

As discussed above, in Kerr spacetime one verifies the absence of a true Hartle-Hawking vacuum, which has led to some different proposals of possible Hartle-Hawking-like vacuum states. The second state we consider in this work is the one introduced by
Frolov and Thorne~\cite{fro3} by using the alternative ``$\eta$ formalism" in order to deal with the quantization of the superradiant modes. This state is invariant under simultaneous $t-\phi$ reversal~\cite{ote}. For more discussions on the Frolov-Thorne vacuum state, we refer the reader the Appendix~\ref{B}.

Let us calculate the rate of variation of atomic energy in the Frolov-Thorne vacuum state. Details concerning the derivation of all the relevant correlation functions of the massless scalar field which appears in Eqs.~(\ref{vfha3}) and~(\ref{rrha3}) can be found in the Appendix~\ref{B}. The contributions of vacuum fluctuations are given by
\bea
\Biggl\langle \frac{d H_A}{dt} \Biggr\rangle_{VF} &=& -\lambda^2\sum_{l,m}\,
\sum_{\nu'}\,{\cal M}(\nu\to\nu')\Delta\nu
\nn\\
&\times&\,\int_{-\Delta\tau}^{\Delta\tau}du\,\Biggl[\int_{0}^{\infty}\,d\omega\coth\left( \frac{\pi\bar{\omega}}{\kappa_{+}} \right)
\nn\\
&\times&\,{\cal P}^{+}_{\omega l m}(r, \theta)\cos(w u) 
\nn\\
&+&\, \int_{0}^{\infty}\,d\bar{\omega} \coth\left( \frac{\pi\bar{\omega}}{\kappa_{+}} \right)
\nn\\
&\times&\,{\cal P}^{-}_{\omega l m}(r, \theta)\cos(w u) \Biggr]\,e^{i\Delta\nu u}.
\eea
By solving the time integrals, one has that
\bea
\Biggl\langle \frac{d H_A}{dt} \Biggr\rangle_{VF} &=& -2\lambda^2\sum_{l,m}\,
\sum_{\nu'}\,{\cal M}(\nu\to\nu')\Delta\nu
\nn\\
&\times&\,\Biggl\{\int_{0}^{\infty}\,d\omega\,\coth\left( \frac{\pi\bar{\omega}}{\kappa_{+}} \right)
\nn\\
&\times&{\cal P}^{+}_{\omega l m}(r, \theta)\, 
{\cal G}_{1}(w, \Delta\nu, \Delta\tau) 
\nn\\
&+&\, \int_{0}^{\infty}\,d\bar{\omega} \coth\left( \frac{\pi\bar{\omega}}{\kappa_{+}} \right)
\nn\\
&\times&\,{\cal P}^{-}_{\omega l m}(r, \theta)\,
{\cal G}_{1}(w, \Delta\nu, \Delta\tau) \Biggr\}.
\eea
In the limit $\Delta\tau \to \infty$, one has:
\bea
\Biggl\langle \frac{d H_A}{dt} \Biggr\rangle_{VF} &=& -\pi\lambda^2\sum_{l}\,
\sum_{\nu'}\,{\cal M}(\nu\to\nu')\widetilde{\Delta\nu}
\nn\\
&\times&\,\theta(\widetilde{\Delta\nu})\,{\cal F}_{l}(1,2,\widetilde{\Delta\nu},r,\theta)
\nn\\
&+&\pi\lambda^2\sum_{l}\,
\sum_{\nu'}\,{\cal M}(\nu\to\nu')|\widetilde{\Delta\nu}|
\nn\\
&\times&\,\theta(-\widetilde{\Delta\nu})\,{\cal F}_{l}(1,2,|\widetilde{\Delta\nu}|,r,\theta),
\eea
where we have defined the quantity 
$$
{\cal F}_{l}(k,n,\widetilde{\Delta\nu},r,\theta) = {\cal F}^{1}_{l}(k,n,\widetilde{\Delta\nu},r,\theta) 
+ {\cal F}^{2}_{l}(k,n,\widetilde{\Delta\nu},r,\theta),
$$ 
together with the definitions given below:\\
\begin{widetext}
\bea
{\cal F}^{1}_{l}(k,n,\widetilde{\Delta\nu},r,\theta) &=& \sum_{m}\left[\left(k + \frac{n}{\exp{\left(\frac{2\pi(\widetilde{\Delta\nu} - m\Omega_{H})}{\kappa_{+}}\right)}-1}\right)
{\cal P}^{+}_{\widetilde{\Delta\nu} l m}(r, \theta) \right.
\nn\\
&+&\, \left. \left(k + \frac{n}{\exp{\left(\frac{2\pi(\widetilde{\Delta\nu} - m\Omega_{H})}{\kappa_{+}}\right)}-1}\right)
{\cal P}^{-}_{\widetilde{\Delta\nu} l m}(r, \theta)\right]\theta(\widetilde{\Delta\nu} - m\Omega_{H}),
\eea
and
\beq
{\cal F}^{2}_{l}(k,n,\widetilde{\Delta\nu},r,\theta) = \sum_{m>0}\left(k + \frac{n}{\exp{\left(\frac{2\pi(\widetilde{\Delta\nu} - m\Omega_{H})}{\kappa_{+}}\right)}-1}\right)
{\cal P}^{+}_{\widetilde{\Delta\nu} l m}(r, \theta)\theta(-\widetilde{\Delta\nu} + m\Omega_{H}). 
\eeq
\end{widetext}
As in the previous case, notice the appearance of the thermal contributions in all terms of the above equation. However, the terms for which $\widetilde{\Delta\nu} < m\Omega_{H}$ (or $|\widetilde{\Delta\nu}| < m\Omega_{H}$) present some interesting consequences. For instance, the thermal term for $|\widetilde{\Delta\nu}| < m\Omega_{H}$ can be rewritten as
\bea
&&1 + \frac{2}{\exp{\left(\frac{2\pi(|\widetilde{\Delta\nu}| - m\Omega_{H})}{\kappa_{+}}\right)}-1} 
\nn\\
&&\,= - \left[1 + \frac{2}{\exp{\left(\frac{2\pi( m\Omega_{H} - |\widetilde{\Delta\nu}|)}{\kappa_{+}}\right)}-1}\right],
\eea
so that for $|\widetilde{\Delta\nu}| < m\Omega_{H}$ and $\widetilde{\Delta\nu} < 0$ vacuum fluctuations leads to rate that obeys: $\langle d H_A/dt \rangle_{VF} < 0$. In other words, the transition $|g\rangle \to |e\rangle$ happens as a negative absorption process. The reader may wonder whether the contribution from radiation reaction would modify this situation. In fact, as in the previous cases one obtains the same results as in the Boulware vacuum state. Hence, when one calculates the total rate, one gets, in the limit $\Delta\tau \to \infty$:
\bea
\Biggl\langle \frac{d H_A}{dt} \Biggr\rangle_{tot} &=& -2\pi\lambda^2\sum_{l}\,
\sum_{\nu'}\,{\cal M}(\nu\to\nu')\widetilde{\Delta\nu}
\nn\\
&\times&\,\theta(\widetilde{\Delta\nu})\,{\cal F}_{l}(1,1,\widetilde{\Delta\nu},r,\theta)
\nn\\
&+& 2\pi\lambda^2\sum_{l}\,
\sum_{\nu'}\,{\cal M}(\nu\to\nu')|\widetilde{\Delta\nu}|
\nn\\
&\times&\,\theta(-\widetilde{\Delta\nu})
\,{\cal F}_{l}(0,1,|\widetilde{\Delta\nu}|,r,\theta).
\eea
The last term of the above expression contains a contribution that represents an absorption process with an overall minus sign. This term would correspond to the transition $|g\rangle \to |e\rangle$. Since the angular velocity is perceived as a chemical potential, one notices that in this situation the chemical potential is greater than the energy gap of this transition. This unusual behavior can be traced back to the following fact~\cite{birrel}. The emission probability depends on $m$, the azimuthal quantum number with respect to the rotation axis, which implies that the emission is asymmetric around the black hole. In general, the thermal factor 
$$
\frac{1}{\exp{\left(\frac{2\pi(|\widetilde{\Delta\nu}| - m\Omega_{H})}{\kappa_{+}}\right)}-1}
$$
is larger for positive $m$ than negative, thus facilitating the emission of quanta with angular momenta oriented towards that of the black hole. However, when $|\widetilde{\Delta\nu}| < m\Omega_{H}$, the thermal factor is negative, as remarked above. Furthermore, even in the limit $M \to \infty$ (temperature goes to zero), the thermal factor remains finite, and equals $-1$, for $|\widetilde{\Delta\nu}| < m\Omega_{H}$ (this is also valid in the extremal situation $a = M$). This negative flux is the consequence of superradiance: The black hole induces stimulated emission. Thus, the absorption probability of the black hole in this situation is negative. In quantum language, this is the result of the Unruh-Starobinskii effect, which has now been understood as the interplay between the two underlying physical effects. So the Frolov-Thorne vacuum state can describe black-hole superradiance in the quantum regime. Finally, as in the previous cases, by using the asymptotic form for the functions ${\cal P}^{\pm}$ given by Eqs.~(\ref{pasymp1}),~(\ref{pasymp2}) one gets the behavior of the rate for $r \to \infty$.

\section{Rate of variation of energy for a stationary atom with zero angular momentum}
\label{rate-stationary}

We now proceed to calculate the rate of variation of energy for an atom following a stationary trajectory but not necessarily with fixed spatial coordinates. In the general stationary case, the atom follows the world line given by $x^{\mu}(\tau) = (t(\tau),r,\theta,\phi(\tau))$, where $r, \theta$ are constants and $\phi(\tau) = \Omega t(\tau)$, $t(\tau) = u^{0}\tau$, with 
$$
u^{0} = \frac{1}{|\Omega^2 g_{\phi\phi} + 2\Omega g_{0\phi} + g_{00}|^{1/2}}
$$ 
and 
$$
\Omega_{-} < \Omega < \Omega_{+}.
$$
Just outside the horizon, the angular velocity of the atom coincides with the angular velocity $\Omega_{H}$ of the black hole. Here we shall consider the special case in which the atom has zero angular momentum. In this situation, the angular velocity of the two-level atom is given by
$$
\Omega = \frac{2 M a r}{(r^2 + a^2)^2 - a^2\Delta\sin^2\theta},
$$
and also $u^{0} = \alpha^{-1}$, with 
$$
\alpha^2 = \frac{\rho^2 \Delta}{\Sigma}.
$$ 
For more details on stationary observers in Kerr spacetime we refere the reader the Appendix~\ref{A}. For simplicity, in this Section we consider large observational times so that the limit $\Delta\tau \to \infty$ is to be understood.

\subsection{The Boulware vacuum state}

Following the same reasoning as above, we start our discussion once again with the Boulware vacuum states. The contributions of vacuum fluctuations read
\bea
\Biggl\langle \frac{d H_A}{dt} \Biggr\rangle_{VF} &=& -\lambda^2\sum_{l,m}\,
\sum_{\nu'}\,{\cal M}(\nu\to\nu')\Delta\nu
\nn\\
&\times&\int_{-\Delta\tau}^{\Delta\tau}du\,\Biggl[\int_{0}^{\infty}\,d\omega\,{\cal P}^{+}_{\omega l m}(r, \theta)\, 
\cos[(\widetilde{w}-\widetilde{m}) u] 
\nn\\
&+&\int_{0}^{\infty}\,d\bar{\omega} {\cal P}^{-}_{\omega l m}(r, \theta)\,
\cos[(\widetilde{w}-\widetilde{m}) u] \Biggr]\,e^{i\Delta\nu u},
\nn\\
\eea
where $\widetilde{w} = \widetilde{w}(\omega) = \omega u^{0}$ and $\widetilde{m} = m\Omega u^{0}$. As above, such an expression was derived for the past modes. Solving the time integrals with $\Delta\tau \to \infty$, one gets (recall that $\Omega \leq \Omega_{H}$)
\bea
\Biggl\langle \frac{d H_A}{dt} \Biggr\rangle_{VF} &=& -\pi\lambda^2\sum_{l}
\sum_{\nu'}\,{\cal M}(\nu\to\nu')\widetilde{\Delta\nu}\,
\nn\\
&\times&\,\theta(\widetilde{\Delta\nu})\,{\cal B}^{+}_{l}(\widetilde{\Delta\nu},r,\theta)
\nn\\
&+&\,\pi\lambda^2\sum_{l}
\sum_{\nu'}\,{\cal M}(\nu\to\nu')|\widetilde{\Delta\nu}|\,
\nn\\
&\times&\,\theta(-\widetilde{\Delta\nu})\,{\cal B}^{+}_{l}(|\widetilde{\Delta\nu}|,r,\theta)
\eea
where
\begin{widetext}
\bea
{\cal B}^{\pm}_{l}(\widetilde{\Delta\nu},r,\theta) &=& \sum_{m, m \neq 0}(\pm)^{\epsilon}
\,{\cal P}^{+}_{(m\Omega + (-1)^{\epsilon} \widetilde{\Delta\nu}) l m}(r, \theta)
\,\theta(m\Omega +(-1)^{\epsilon} \widetilde{\Delta\nu})
\nn\\
&+&\,  \sum_{m, m \neq 0}(\pm)^{1-\epsilon}\,{\cal P}^{-}_{(m\Omega - (-1)^{\epsilon} \widetilde{\Delta\nu}) l m}(r, \theta)
\,\theta(m(\Omega-\Omega_{H}) - (-1)^{\epsilon} \widetilde{\Delta\nu})
+ \sum_{m}{\cal P}^{\sgn(m)}_{(m\Omega + \widetilde{\Delta\nu}) l m}(r, \theta),
\eea
\end{widetext}
with $\epsilon = \theta(m)$. Notice the contribution of a term that is dependent of a relation between the angular velocity of the event horizon and the angular velocity of the qubit. In addition, in the above expression we considered that 
\bea
{\cal P}^{\sgn(m)}_{\omega l m} &=& {\cal P}^{+}_{\omega l m},\,\,\,m > 0
\nn\\
{\cal P}^{\sgn(m)}_{\omega l m} &=& {\cal P}^{-}_{\omega l m},\,\,\,m < 0
\nn\\
{\cal P}^{0}_{\omega l 0} &=& {\cal P}^{+}_{\omega l 0} + {\cal P}^{-}_{\omega l 0}.
\eea
Observe that, by assessing only the vacuum-fluctuation contribution, both spontaneous excitation and deexcitation
occur with equal magnitude, a result similar to the analogous case treated previously. 

The contribution from radiation reaction reads
\bea
\Biggl\langle \frac{d H_A}{dt} \Biggr\rangle_{RR} &=& i\lambda^2\sum_{l,m}\,\sum_{\nu'}\,
{\cal M}(\nu\to\nu')\Delta\nu
\nn\\
&\times&\int_{-\Delta\tau}^{\Delta\tau}du
\Biggl[\int_{0}^{\infty}\,d\omega\,{\cal P}^{+}_{\omega l m}(r, \theta)\,\sin[(\widetilde{w}-\widetilde{m}) u] 
\nn\\
&+&\, \int_{0}^{\infty}\,d\bar{\omega} {\cal P}^{-}_{\omega l m}(r, \theta)\,\sin[(\widetilde{w}-\widetilde{m}) u] \Biggr]\,e^{i\Delta\nu u}.
\nn\\
\label{rr-boul3}
\eea
The above expression is valid for both past and future modes. The time integrals can be easily solved in the limit $\Delta\tau \to \infty$ and the result is
\bea
\Biggl\langle \frac{d H_A}{dt} \Biggr\rangle_{RR} &=& -\pi\lambda^2\sum_{l}
\sum_{\nu'}\,{\cal M}(\nu\to\nu')\widetilde{\Delta\nu}
\nn\\
&\times&\,\theta(\widetilde{\Delta\nu})\,{\cal B}^{-}_{l}(\widetilde{\Delta\nu},r,\theta)
\nn\\
&-&\,\pi\lambda^2\sum_{l}
\sum_{\nu'}\,{\cal M}(\nu\to\nu')|\widetilde{\Delta\nu}|
\nn\\
&\times&\,\theta(-\widetilde{\Delta\nu})\,{\cal B}^{-}_{l}(|\widetilde{\Delta\nu}|,r,\theta).
\eea
By observing the above expression, one notes that, depending on the relation between the energy gap, $\Omega$ and $\Omega_{H}$, the effect of radiation reaction may lead to a gain of atomic energy. This is a different situation in comparison with the analogous one in the case of a static atom and leads to an interesting result. To see this, consider the total rate for $\widetilde{\Delta\nu} < 0$, i.e., the transition $|g\rangle \to |e\rangle$. After summing up the associated contributions of vacuum fluctuations and radiation reaction, one finds:
\bea
&&\Biggl\langle \frac{d H_A}{dt} \Biggr\rangle_{tot} = 2\pi\lambda^2\,\widetilde{\omega}_0
\nn\\
&\times&\,\left[\sum_{l,m>0}{\cal P}^{+}_{(m\Omega - \widetilde{\omega}_0) l m}(r, \theta)\,
\theta(m\Omega - \widetilde{\omega}_0) \right.
\nn\\
&+& \left. \sum_{l,m<0}\,{\cal P}^{-}_{(m\Omega - \widetilde{\omega}_0) l m}(r, \theta) \,
\theta(m(\Omega- \Omega_{H}) -\widetilde{\omega}_0)\right].
\eea
There is no perfect balance between vacuum fluctuations and radiation reaction, and, accordingly, transitions to the excited state become possible even in the vacuum for large observational times. Clearly this happens when the energy gap between quantum states of the atom satisfies the superradiant condition. Being more specific, in the ZAMOs perspective, the Boulware vacuum state contains an outward flux of particles obeying the superradiant condition. Taken in connection with the associated outcome for static atoms, this clarifies the earlier result that ground-state static atoms are stable in this case, even though, as mentioned, the past Boulware vacuum state does not correspond to the concept of a most empty state at infinity. Since $\widetilde{\omega}_0 < m\Omega$ with $m > 0$ (or $\widetilde{\omega}_0 < m(\Omega - \Omega_{H})$, $m < 0$), this spontaneous excitation has a nontrivial relationship with the Unruh-Starobinskii effect. Yet, notice that, as $r \to r_{+}$, $\Omega \to \Omega_{H}$ and, for $m\Omega_{H} < \widetilde{\omega}_0$, the effect ceases to exist, and the balance between vacuum fluctuations and radiation reaction is restored. In turn, ${\cal P}^{-}$ is vanishingly small as $r \to \infty$ and hence this effect is also suppressed at spatial infinity. In any case, it is not clear whether such results could be understood in terms of a curved-space generalization of a rotating quantum vacuum in flat spacetime~\cite{rot3,rot4}. 

\subsection{The Unruh vacuum state}

Now let us derive the contributions to the rate of variation of atomic energy in the case of a gravitational collapse of a rotating massive body. The contributions of vacuum fluctuations are given by
\bea
\Biggl\langle \frac{d H_A}{dt} \Biggr\rangle_{VF} &=& -\lambda^2\sum_{l,m}\,
\sum_{\nu'}\,{\cal M}(\nu\to\nu')\Delta\nu
\nn\\
&\times&\int_{-\Delta\tau}^{\Delta\tau}du\Biggl[\int_{0}^{\infty}\,d\omega\,{\cal P}^{+}_{\omega l m}(r, \theta)
\cos[(\widetilde{w}-\widetilde{m}) u]
\nn\\
&+&\, \int_{0}^{\infty}\,d\bar{\omega} \coth\left( \frac{\pi\bar{\omega}}{\kappa_{+}} \right)
\nn\\
&\times&\,{\cal P}^{-}_{\omega l m}(r, \theta)\,\cos[(\widetilde{w}-\widetilde{m}) u] \Biggr]\,e^{i\Delta\nu u}.
\eea
Solving the time integrals, one gets, in the limit $\Delta\tau \to \infty$
\bea
\Biggl\langle \frac{d H_A}{dt} \Biggr\rangle_{VF} &=& -\pi\lambda^2\sum_{l}
\sum_{\nu'}\,{\cal M}(\nu\to\nu')\widetilde{\Delta\nu}
\nn\\
&\times&\,\theta(\widetilde{\Delta\nu})\,{\cal V}_{l}(\widetilde{\Delta\nu},r,\theta)
\nn\\
&+&\,\pi\lambda^2\sum_{l}
\sum_{\nu'}\,{\cal M}(\nu\to\nu')|\widetilde{\Delta\nu}|
\nn\\
&\times&\,\theta(-\widetilde{\Delta\nu})\,{\cal V}_{l}(|\widetilde{\Delta\nu}|,r,\theta),
\eea
where we have introduced the quantity:
\begin{widetext}
\bea
{\cal V}_{l}(\widetilde{\Delta\nu},r,\theta) &=& \sum_{m, m \neq 0}
{\cal P}^{+}_{(m\Omega + (-1)^{\epsilon}\widetilde{\Delta\nu}) l m}(r, \theta)
\,\theta(m\Omega + (-1)^{\epsilon}\widetilde{\Delta\nu})
\nn\\
&& +  \sum_{m, m \neq 0}\,\left[1 + \frac{2}{\exp{\left[\frac{2\pi[-(-1)^{\epsilon}\widetilde{\Delta\nu} + m(\Omega-\Omega_{H})]}
{\kappa_{+}} \right]} - 1}\right]\,
{\cal P}^{-}_{(m\Omega -(-1)^{\epsilon} \widetilde{\Delta\nu}) l m}(r, \theta)
\,\theta(m(\Omega-\Omega_{H}) -(-1)^{\epsilon} \widetilde{\Delta\nu}) 
\nn\\
&+&\, \sum_{m}\,\left[{\cal P}^{\sgn(m)}_{(m\Omega + \widetilde{\Delta\nu}) l m}(r, \theta) 
+ \frac{2\,\theta(-m)\,{\cal P}^{\Sgn(m)}_{(m\Omega + \widetilde{\Delta\nu}) l m}(r, \theta)}{\exp{\left[\frac{2\pi[\widetilde{\Delta\nu} - m(\Omega_{H}-\Omega)]}
{\kappa_{+}} \right]} - 1}\right].
\eea
(In the above we considered that $\theta(x) = 1/2$ for $x = 0$ and ${\cal P}^{\Sgn(m)}_{\omega l m} 
= 2\,{\cal P}^{-}_{\omega l m}$, for $m = 0$.) Notice again the presence of thermal terms that enhance the atomic radiative processes concerning only the vacuum fluctuations.

As for the contribution from radiation reaction, it is easy to prove from the correlation functions presented in the Appendix~\ref{B} that we obtain the same expression as in the previous case, namely Eq.~(\ref{rr-boul3}). Hence the sum of both contributions leads to the total rate; explicitly one finds, for $\widetilde{\Delta\nu} > 0$:
\bea
&&\Biggl\langle \frac{d H_A}{dt} \Biggr\rangle_{tot} = -2\pi\lambda^2\,\widetilde{\omega}_0\,
\left\{\sum_{l,m<0}\,{\cal P}^{+}_{(m\Omega + \widetilde{\omega}_0) l m}(r, \theta)\,
\theta(m\Omega + \widetilde{\omega}_0) \right.
\nn\\
&&+\, \left. \sum_{l,m,m \neq 0}\left[\theta(m)+\frac{1}{\exp{\left[\frac{2\pi[-(-1)^{\epsilon}\widetilde{\omega}_0 + m(\Omega-\Omega_{H})]}{\kappa_{+}} \right]} - 1}\right]\,
{\cal P}^{-}_{(m\Omega - (-1)^{\epsilon}\widetilde{\omega}_0) l m}(r, \theta)\,
\theta(m(\Omega-\Omega_{H}) - (-1)^{\epsilon}\widetilde{\omega}_0) \right.
\nn\\
&&+\, \left. \sum_{l,m}\,\left[{\cal P}^{\sgn(m)}_{(m\Omega + \widetilde{\omega}_0) l m}(r, \theta) + 
\frac{\theta(-m)\,{\cal P}^{\Sgn(m)}_{(m\Omega + \widetilde{\omega}_0) l m}(r, \theta)}{\exp{\left[\frac{2\pi[\widetilde{\omega}_0 - m(\Omega_{H}-\Omega)]}
{\kappa_{+}} \right]} - 1}\right]\right\},
\label{eq-61}
\eea
and, for $\widetilde{\Delta\nu} < 0$:
\bea
&&\Biggl\langle \frac{d H_A}{dt} \Biggr\rangle_{tot} = 2\pi\lambda^2\,\widetilde{\omega}_0
\left\{\sum_{l,m>0}{\cal P}^{+}_{(m\Omega - \widetilde{\omega}_0) l m}(r, \theta)\,\theta(m\Omega - \widetilde{\omega}_0) 
\right.
\nn\\
&&\,+ \left. \sum_{l,m,m \neq 0}\left[\theta(-m)+\frac{1}{\exp{\left[\frac{2\pi[-(-1)^{\epsilon}\widetilde{\omega}_0 + m(\Omega-\Omega_{H})]}{\kappa_{+}} \right]} - 1}\right]\,
{\cal P}^{-}_{(m\Omega - (-1)^{\epsilon}\widetilde{\omega}_0) l m}(r, \theta)\,
\theta(m(\Omega-\Omega_{H}) - (-1)^{\epsilon}\widetilde{\omega}_0)  \right.
\nn\\
&&+\, \left. \sum_{l,m<0}\,\left[\frac{1}{\exp{\left[\frac{2\pi[\widetilde{\omega}_0 - m(\Omega_{H}-\Omega)]}
{\kappa_{+}} \right]} - 1}\right]{\cal P}^{-}_{(m\Omega + \widetilde{\omega}_0) l m}(r, \theta)\right\}.
\eea
We observe again the appearance of thermal terms associated with the Hawking effect. On the other hand, one verifies the emergence of unusual thermal terms, see for instance the second line in Eq.~(\ref{eq-61}) ($\langle d H_A/dt \rangle_{VF} < 0$). An analogous result holds for $\langle d H_A/dt \rangle_{VF} > 0$. Since this phenomenon ensues whenever $m(\Omega - \Omega_{H}) > \widetilde{\omega}_0$, with negative $m$, based on the discussions above one interprets the ocurrence of such terms as a consequence of the Unruh-Starobinskii effect. Thence one concludes that the Hawking effect and the Unruh-Starobinskii effect are competing processes concerning radiative processes of stationary atoms near a massive spinning star undergoing gravitational collapse to form a rotating black hole. This is the physical content of the (past) Unruh vacuum state in the ZAMOs perspective.

In order to get a better insight on the above results, let us investigate the behavior of the rate in the two asymptotic regions of interest. For $r \to \infty$, Eqs.~(\ref{pasymp1}) and~(\ref{pasymp2}) reveal that the rate is intensively suppressed at spatial infinity; yet, depending on the relationship between $\Omega$, $\Omega_{H}$ and the energy gap, such a suppression may be reduced or even precluded. This happens whenever the energy gap $\widetilde{\omega}_0$ takes values close enough to $m\Omega$ or $m(\Omega_{H} - \Omega)$. On the other hand, inside the ergosphere and near the horizon, $\Omega \to \Omega_{H}$, and the terms obeying the superradiant condition do not contribute to the rate. In turn, the excitation rate is larger at ${\cal H}^{+}$ in comparison with ${\cal H}^{-}$.

\subsection{The Candelas-Chrzanowski-Howard vacuum state}

Now we turn to the considerations associated with the candidates of a Hartle-Hawking vacuum states. As above, we begin with the Candelas-Chrzanowski-Howard vacuum state. The contributions of vacuum fluctuations are given by
\bea
\Biggl\langle \frac{d H_A}{dt} \Biggr\rangle_{VF} &=& -\lambda^2\sum_{l,m}\,
\sum_{\nu'}\,{\cal M}(\nu\to\nu')\Delta\nu
\nn\\
&\times&\,\int_{-\Delta\tau}^{\Delta\tau}du\,\Biggl[\int_{0}^{\infty}\,d\omega\coth\left( \frac{\pi\omega}{\kappa_{+}} \right)
\,{\cal P}^{+}_{\omega l m}(r, \theta)\, \cos[(\widetilde{w}-\widetilde{m}) u] 
\nn\\
&+&\, \int_{0}^{\infty}\,d\bar{\omega} \coth\left( \frac{\pi\bar{\omega}}{\kappa_{+}} \right)
\,{\cal P}^{-}_{\omega l m}(r, \theta)\,\cos[(\widetilde{w}-\widetilde{m}) u] \Biggr]\,e^{i\Delta\nu u}.
\eea
Solving the time integrals, one gets, in the limit $\Delta\tau \to \infty$
\bea
\Biggl\langle \frac{d H_A}{dt} \Biggr\rangle_{VF} &=& -\pi\lambda^2\sum_{l}
\sum_{\nu'}\,{\cal M}(\nu\to\nu')\widetilde{\Delta\nu}
\,\theta(\widetilde{\Delta\nu})\,{\cal D}_{l}(\widetilde{\Delta\nu},r,\theta)
\nn\\
&+&\,\pi\lambda^2\sum_{l}
\sum_{\nu'}\,{\cal M}(\nu\to\nu')|\widetilde{\Delta\nu}|
\,\theta(-\widetilde{\Delta\nu})\,{\cal D}_{l}(|\widetilde{\Delta\nu}|,r,\theta),
\eea
where
\bea
{\cal D}_{l}(\widetilde{\Delta\nu},r,\theta) &=& \sum_{m,m \neq 0}\left[1 + \frac{2}{\exp{\left[\frac{2\pi[(-1)^{\epsilon}\widetilde{\Delta\nu} +  m\Omega]}
{\kappa_{+}} \right]} - 1}\right]
{\cal P}^{+}_{(m\Omega +(-1)^{\epsilon} \widetilde{\Delta\nu}) l m}(r, \theta)
\,\theta(m\Omega +(-1)^{\epsilon} \widetilde{\Delta\nu}) 
\nn\\
&+&\,  \sum_{m,m \neq 0}\left[1 + \frac{2}{\exp{\left[\frac{2\pi[-(-1)^{\epsilon}\widetilde{\Delta\nu} + m(\Omega-\Omega_{H})]}
{\kappa_{+}} \right]} - 1}\right]\,
{\cal P}^{-}_{(m\Omega - (-1)^{\epsilon}\widetilde{\Delta\nu}) l m}(r, \theta)\,
\theta(m(\Omega-\Omega_{H}) - (-1)^{\epsilon}\widetilde{\Delta\nu}) 
\nn\\
&+&\,  \sum_{m}\left[1 + \frac{2\,\theta(m)}{\exp{\left[\frac{2\pi[\widetilde{\Delta\nu} + m\Omega]}
{\kappa_{+}} \right]} - 1}+ \frac{2\,\theta(-m)}{\exp{\left[\frac{2\pi[\widetilde{\Delta\nu} - m(\Omega_{H}-\Omega)]}
{\kappa_{+}} \right]} - 1}\right]\,{\cal P}^{\sgn(m)}_{(m\Omega + \widetilde{\Delta\nu}) l m}(r, \theta).
\eea
Now all terms are multiplied by a Planck factor, in contrast to the case considered previously.

Concerning the contribution from radiation reaction, as in the previous case we obtain the same results as in the Boulware vacuum states. Hence, the total rate is given by, with $\widetilde{\Delta\nu} > 0$:
\bea
\Biggl\langle \frac{d H_A}{dt} \Biggr\rangle_{tot} &=& -2\pi\lambda^2\,\widetilde{\omega}_0
\left\{\sum_{l,m,m \neq 0}\left[\theta(-m) + \frac{1}{\exp{\left[\frac{2\pi[(-1)^{\epsilon}\widetilde{\omega}_0 +  m\Omega]}
{\kappa_{+}} \right]} - 1}\right]
{\cal P}^{+}_{(m\Omega +(-1)^{\epsilon} \widetilde{\omega}_0) l m}(r, \theta)
\,\theta(m\Omega +(-1)^{\epsilon} \widetilde{\omega}_0) \right.
\nn\\
&+&\, \left. \sum_{l,m,m \neq 0}\left[\theta(m) + \frac{1}{\exp{\left[\frac{2\pi[-(-1)^{\epsilon}\widetilde{\omega}_0 + m(\Omega-\Omega_{H})]}
{\kappa_{+}} \right]} - 1}\right]\,
{\cal P}^{-}_{(m\Omega - (-1)^{\epsilon}\widetilde{\omega}_0) l m}(r, \theta)\,
\theta(m(\Omega-\Omega_{H}) - (-1)^{\epsilon}\widetilde{\omega}_0) \right.
\nn\\
&+&\, \left. \sum_{l,m}\left[1 + \frac{\theta(m)}{\exp{\left[\frac{2\pi[\widetilde{\omega}_0 + m\Omega]}
{\kappa_{+}} \right]} - 1} + \frac{\theta(-m)}{\exp{\left[\frac{2\pi[\widetilde{\omega}_0 + m(\Omega-\Omega_{H})]}
{\kappa_{+}} \right]} - 1}\right]\,{\cal P}^{\sgn(m)}_{(m\Omega + \widetilde{\omega}_0) l m}(r, \theta)\right\}.
\eea
For $\widetilde{\Delta\nu} < 0$, one gets:
\bea
\Biggl\langle \frac{d H_A}{dt} \Biggr\rangle_{tot} &=& 2\pi\lambda^2\,\widetilde{\omega}_0
\left\{\sum_{l,m,m \neq 0}\left[\theta(m) + \frac{1}{\exp{\left[\frac{2\pi[(-1)^{\epsilon}\widetilde{\omega}_0 +  m\Omega]}
{\kappa_{+}} \right]} - 1}\right]
{\cal P}^{+}_{(m\Omega +(-1)^{\epsilon} \widetilde{\omega}_0) l m}(r, \theta)
\,\theta(m\Omega + (-1)^{\epsilon} \widetilde{\omega}_0) \right.
\nn\\
&+&\, \left. \sum_{l,m,m \neq 0}\left[\theta(-m) + \frac{1}{\exp{\left[\frac{2\pi[-(-1)^{\epsilon}\widetilde{\omega}_0 + m(\Omega-\Omega_{H})]}{\kappa_{+}} \right]} - 1}\right]\,
{\cal P}^{-}_{(m\Omega - (-1)^{\epsilon}\widetilde{\omega}_0) l m}(r, \theta)\,
\theta(m(\Omega-\Omega_{H}) - (-1)^{\epsilon}\widetilde{\omega}_0) \right.
\nn\\
&+&\, \left. \sum_{l,m}\left[\frac{\theta(m)}{\exp{\left[\frac{2\pi[\widetilde{\omega}_0 + m\Omega]}
{\kappa_{+}} \right]} - 1} + \frac{\theta(-m)}{\exp{\left[\frac{2\pi[\widetilde{\omega}_0 + m(\Omega-\Omega_{H})]}
{\kappa_{+}} \right]} - 1}\right]\,{\cal P}^{\sgn(m)}_{(m\Omega + \widetilde{\omega}_0) l m}(r, \theta)\right\}.
\eea
Let us discuss the physical meaning of the the Candelas-Chrzanowski-Howard vacuum state under a ZAMO point of view. One verifies the appearance of thermal contributions even when the atomic energy gap satisfies the superradiant condition. However, here we identify the ocurrence of such terms when $m(\Omega - \Omega_{H}) > \widetilde{\omega}_0$ (negative $m$) but also when $m\Omega > \widetilde{\omega}_0$ (positive $m$), in contrast with the Unruh Vacuum state. This is the result of the fact that the Candelas-Chrzanowski-Howard vacuum state does not respect the simultaneous $t-\phi$ reversal invariance of Kerr spacetime, as discussed above.

For completeness, let us analyze the rate in the asymptotic regions. By using Eqs.~(\ref{pasymp1}) and~(\ref{pasymp2}), one notes that the rate is suppressed at spatial infinity. However, when the energy gap $\widetilde{\omega}_0$ takes values close enough to $m\Omega$ or $m(\Omega_{H} - \Omega)$, the rate undergoes a violent increase. On the other hand, near the horizon, $\Omega \to \Omega_{H}$, but, unlike the previous case, the terms obeying the superradiant condition do contribute to the rate. In addition, as above the excitation rate is larger at ${\cal H}^{+}$ in comparison with ${\cal H}^{-}$.

\subsection{The Frolov-Thorne vacuum state}

To conclude our discussions, we finally turn our attentions to the Frolov-Thorne vacuum state. The contributions of vacuum fluctuations are given by
\bea
\Biggl\langle \frac{d H_A}{dt} \Biggr\rangle_{VF} &=& -\lambda^2\sum_{l,m}\,
\sum_{\nu'}\,{\cal M}(\nu\to\nu')\Delta\nu
\nn\\
&\times&\,\int_{-\Delta\tau}^{\Delta\tau}du\,\Biggl[\int_{0}^{\infty}\,d\omega\coth\left( \frac{\pi\bar{\omega}}{\kappa_{+}} \right)
\,{\cal P}^{+}_{\omega l m}(r, \theta)\, \cos[(\widetilde{w}-\widetilde{m}) u] 
\nn\\
&+&\, \int_{0}^{\infty}\,d\bar{\omega} \coth\left( \frac{\pi\bar{\omega}}{\kappa_{+}} \right)
\,{\cal P}^{-}_{\omega l m}(r, \theta)\,\cos[(\widetilde{w}-\widetilde{m}) u]\Biggr]\,e^{i\Delta\nu u}.
\eea
Solving the time integrals, one gets, in the limit $\Delta\tau \to \infty$
\bea
\Biggl\langle \frac{d H_A}{dt} \Biggr\rangle_{VF} &=& -\pi\lambda^2\sum_{l}
\sum_{\nu'}\,{\cal M}(\nu\to\nu')\widetilde{\Delta\nu}
\,\theta(\widetilde{\Delta\nu})\,{\cal Q}_{l}(\widetilde{\Delta\nu},r,\theta)
\nn\\
&+&\,\pi\lambda^2\sum_{l}
\sum_{\nu'}\,{\cal M}(\nu\to\nu')|\widetilde{\Delta\nu}|
\,\theta(-\widetilde{\Delta\nu})\,{\cal Q}_{l}(|\widetilde{\Delta\nu}|,r,\theta),
\eea
where
\bea
{\cal Q}_{l}(\widetilde{\Delta\nu},r,\theta) &=& \sum_{m,m \neq 0}\,(-1)^{\epsilon}\,
\left[1 + \frac{2}{\exp{\left[\frac{2\pi[\widetilde{\Delta\nu} - \sgn(m)\,m\,(\Omega-\Omega_{H})]}
{\kappa_{+}} \right]} - 1}\right]
{\cal P}^{+}_{(m\Omega +(-1)^{\epsilon} \widetilde{\Delta\nu}) l m}(r, \theta)
\,\theta(m\Omega +(-1)^{\epsilon} \widetilde{\Delta\nu}) 
\nn\\
&+&\,  \sum_{m,m \neq 0}\left[1 + \frac{2}{\exp{\left[\frac{2\pi[-(-1)^{\epsilon}\widetilde{\Delta\nu} + m(\Omega-\Omega_{H})]}
{\kappa_{+}} \right]} - 1}\right]\,
{\cal P}^{-}_{(m\Omega - (-1)^{\epsilon}\widetilde{\Delta\nu}) l m}(r, \theta)\,
\theta(m(\Omega-\Omega_{H}) - (-1)^{\epsilon}\widetilde{\Delta\nu}) 
\nn\\
&+&\,  \sum_{m}\left[1 + \frac{2}{\exp{\left[\frac{2\pi[\widetilde{\Delta\nu} + m(\Omega-\Omega_{H})]}
{\kappa_{+}} \right]} - 1}\right]
{\cal P}^{\sgn(m)}_{(m\Omega + \widetilde{\Delta\nu}) l m}(r, \theta).
\eea
In turn, again the contribution from radiation reaction is given by the same results as in the Boulware vacuum states. Hence the total rate reads, with $\widetilde{\Delta\nu} > 0$
\bea
\Biggl\langle \frac{d H_A}{dt} \Biggr\rangle_{tot} &=& -2\pi\lambda^2\,\widetilde{\omega}_0
\left\{\sum_{l,m,m \neq 0}\,(-1)^{\epsilon}\,
\left[1 + \frac{1}{\exp{\left[\frac{2\pi[\widetilde{\omega}_0 - \sgn(m)\,m\,(\Omega-\Omega_{H})]}
{\kappa_{+}} \right]} - 1}\right]
{\cal P}^{+}_{(m\Omega +(-1)^{\epsilon} \widetilde{\omega}_0) l m}(r, \theta)
\,\theta(m\Omega +(-1)^{\epsilon} \widetilde{\omega}_0) \right.
\nn\\
&+&\, \left. \sum_{l,m,m \neq 0}
\left[\theta(m) + \frac{1}{\exp{\left[\frac{2\pi[-(-1)^{\epsilon}\widetilde{\omega}_0 + m(\Omega-\Omega_{H})]}
{\kappa_{+}} \right]} - 1}\right]\,
{\cal P}^{-}_{(m\Omega - (-1)^{\epsilon}\widetilde{\omega}_0) l m}(r, \theta)\,
\theta(m(\Omega-\Omega_{H}) - (-1)^{\epsilon}\widetilde{\omega}_0) \right.
\nn\\
&+&\, \left. \sum_{l,m}\left[1 + \frac{1}{\exp{\left[\frac{2\pi[\widetilde{\omega}_0 + m(\Omega-\Omega_{H})]}
{\kappa_{+}} \right]} - 1}\right]
{\cal P}^{\sgn(m)}_{(m\Omega + \widetilde{\omega}_0) l m}(r, \theta)\right\},
\eea
whereas for $\widetilde{\Delta\nu} < 0$ one finds:
\bea
\Biggl\langle \frac{d H_A}{dt} \Biggr\rangle_{tot} &=& 2\pi\lambda^2\,\widetilde{\omega}_0
\left\{\sum_{l,m,m \neq 0}\,(-1)^{\epsilon}\,
\left[\frac{1}{\exp{\left[\frac{2\pi[\widetilde{\omega}_0 - \sgn(m)\,m\,(\Omega-\Omega_{H})]}
{\kappa_{+}} \right]} - 1}\right]
{\cal P}^{+}_{(m\Omega +(-1)^{\epsilon} \widetilde{\omega}_0) l m}(r, \theta)
\,\theta(m\Omega + (-1)^{\epsilon} \widetilde{\omega}_0) \right.
\nn\\
&+&\, \left.  \sum_{l,m,m \neq 0}\left[\theta(-m) + \frac{1}{\exp{\left[\frac{2\pi[-(-1)^{\epsilon}\widetilde{\omega}_0 + m(\Omega-\Omega_{H})]}{\kappa_{+}} \right]} - 1}\right]\,
{\cal P}^{-}_{(m\Omega - (-1)^{\epsilon}\widetilde{\omega}_0) l m}(r, \theta)\,
\theta(m(\Omega-\Omega_{H}) - (-1)^{\epsilon}\widetilde{\omega}_0) \right.
\nn\\
&+&\, \left. \sum_{l,m}\left[\frac{1}{\exp{\left[\frac{2\pi[\widetilde{\omega}_0 + m(\Omega-\Omega_{H})]}
{\kappa_{+}} \right]} - 1}\right]\,{\cal P}^{\sgn(m)}_{(m\Omega + \widetilde{\omega}_0) l m}(r, \theta)\right\}.
\eea
\end{widetext}
We observe again the appearance of thermal terms which one may interpret as being the case when the black hole is immersed in a bath of thermal radiation at the Hawking temperature. In addition, terms related to the Unruh-Starobinskii process also emerges in the above expressions whenever the energy gap satisfies the superradiant condition. Within the physical nature of the definition of the Frolov-Thorne vacuum state, we also observe a negative term associated with absorption process and a positive term related to emission process -- again this is connected with the Unruh-Starobinskii effect. Both phenomena have a nontrivial relationship with the spontaneous excitation of atoms in the ground state. Furthermore, in the asymptotic regions, we obtain similar conclusions as in the previous case: In particular, at the resonant frequencies $m(\Omega_{H} - \Omega) \gtrsim \widetilde{\omega}_0$, the rate is arbitrarily large.

\section{Conclusions and Perspectives}
\label{conclude}

Based upon the formalism introduced by Dalibard, Dupont-Roc, and Cohen-Tannoudji, we have discussed the distinct effects of vacuum fluctuations and radiation reaction on radiative processes of a two-level atom in Kerr spacetime. We considered the qubit coupled to a massless quantum scalar field. The overall picture is the following. The rate of change of the atomic energy is very small when the two-level atom is placed far away of the horizon for both trajectories considered in this paper. For the case in which the qubit follows a stationary path with zero angular momentum, we have shown that the excitation rate is larger at ${\cal H}^{+}$ in comparison with ${\cal H}^{-}$. Depending on the relative values of the energy gap and the angular velocity of the event horizon, radiative processes of stationary qubits with zero angular momentum can be highly magnified near the horizon. We also have demonstrated in which situations black-hole superradiance is connected with spontaneous excitation of atoms. 

The purpose of this work was to search for an interpretation of the Hawking and the Unruh-Starobinskii effects in a rotating black hole as the result of the interplay between vacuum fluctuations and radiation reaction. We have computed the contributions to the variation of the mean atomic energy for two cases, namely a static atom and a stationary atom with zero angular momentum. For the field prepared in the Boulware vacuum state, we have shown that, for a static atom in the ground state, the perfect balance between vacuum fluctuations and radiation reaction avoids the possibility of spontaneous excitation, although the Boulware vacuum state in Kerr spacetime does not agree with the concept of a state which is most empty at infinity. This physical interpretation resembles the usually given for a static atom interacting with a quantum field in the Minkowski vacuum. 

For the other vacuum states considered here, the balance between vacuum fluctuations and radiation reaction is disturbed in virtue of the presence of thermal radiation and static atoms may get excited due to their interaction with a radiation field as a consequence of the Hawking effect. The physical meaning of each one is, however, different. The (past) Unruh vacuum state describes a rotating black hole with an outgoing flux of thermal radiation which does not obey the usual superradiant condition, whereas the Candelas-Chrzanowski-Howard vacuum state represents at ${\cal J}^{-}$ a rotating black hole in thermal equilibrium with an infinite sea of black-body radiation; it is a thermal state and the thermal properties are not related with the existence of any specific requisition associated with the superradiant condition. However, this latter vacuum state does not respect the simultaneous $t-\phi$ reversal invariance of Kerr spacetime, as shown above. 

In turn, when the field is prepared in the Frolov-Thorne vacuum state, we gave evidences in favour of two possible interpretations. Such a state can describe a rotating black hole immersed in a bath of thermal radiation at the Hawking temperature; nevertheless we have proved in this case that spontaneous excitation from the ground state for static atoms is also related to the Unruh-Starobinskii effect (the quantum equivalent of superradiance). The requirement which supports this view is given whenever the energy gap obeys a superradiant condition. Hence the Frolov-Thorne vacuum state describes black-hole superradiance in the quantum regime. However, this vacuum state fails to be regular almost everywhere as demonstrated in Ref.~\cite{ote}. 

On the other hand, in the ZAMOs perspective the above physical interpretation may change drastically. For instance, black-hole superradiance also implies that atoms with nonvanishing angular velocity (but zero angular momentum) coupled with quantum fields in Boulware vacuum states may undergo transitions to higher levels, again within a superradiant circumstance. This result enlightens the fact that the (past) Boulware vacuum state does not correspond to the notion of a most empty state at infinity, yet ground-state static atoms are stable in this situation. In this ZAMO case, we also have shown in what situations the Hawking effect and the Unruh-Starobinskii effect are competing processes in the spontaneous emission and absorption processes undergone by atoms. Both effects have been traced back to the interplay between the two underlying physical effects, vacuum flucutations and radiation reaction. In particular, the (past) Unruh vacuum state does not describe the Unruh-Starobinskii effect inside the ergosphere, whereas this phenomenon is verified in this region for the Candelas-Chrzanowski-Howard vacuum state.

We believe that the results discussed here are pivotal for the investigations of radiative processes of atoms in a rotating black-hole background. For instance, it is known that the Hawking effect of the Kerr spacetime can be understood as the the manifestation of thermalization phenomena via open quantum system approach~\cite{xian}. Such results can be combined with the outcomes deduced here in the direction of a thorough comprehension of the relation between the Hawking effect and the spontaneous excitation of atoms in Kerr spacetime. In turn, the results presented in this paper may also have an impact in the studies of black-hole superradiance and its quantum counterpart, the Unruh-Starobinskii effect. In addition to the important topics remarked in the text, black-hole superradiance has also a remarkable relationship with phase transitions between spinning or charged black objects in asymptotically AdS spacetime~\cite{cardoso2,dias2,dias}. Such important investigations may benefit from a sistematic examination of radiative processes near the event horizon of a rotating black hole. We also mention that astrophysical processes underlying the emission of jets from accretion disks surrounding black holes may have some connection with the outcomes of the present work. For instance, in Ref.~\cite{koba}, the black-hole superradiance of electromagnetic waves emitted from the disk surface was considered in detail. In summary, a framework in which vacuum fluctuations and radiation-reaction effect have been clearly identified and quantitatively analyzed may contribute to an accurate and deeper understanding of such results. In any case, we understand that studies involving rotating black holes may contribute to attain more insights on the quantum-gravity structure of the physical vacuum.

As discussed above, this work indicates that it is possible to put together concepts from black-hole superradiance, relativistic jets and quantum information theory within the DDC approach. Concerning this last subject, studies reported recently confirm the relevance of the field of relativistic quantum information as a vivid research program~\cite{peres,cqg,fuentes}. Since, as already quoted above, the DDC formalism was sucessfully employed in order to study quantum entanglement~\cite{ng1,ng2,ng3}, a natural extension of such papers is to include the investigation of entanglement generation between atoms in Kerr spacetime resorting to the same method. On the other hand, one could envisage the same situation within a more standard formalism such as the traditional master equation approach. In turn, we also believe that the outcomes presented here may be pertinent to the studies of Casimir-Polder interaction between two atoms near the event horizon of a rotating black hole. Indeed, recently the DDC formalism was employed to study the Casimir-Polder forces between two uniformly accelerated atoms~\cite{marino}. Such subjects are under investigation and results will be reported elsewhere.

\section*{acknowlegements}

Work supported by Conselho Nacional de Desenvolvimento Cientifico e Tecnol{\'o}gico do Brasil (CNPq).

\appendix

\section{Observers in Kerr spacetime}
\label{A}

 In the Kerr spacetime there are three families of observers of interest~\cite{poisson}. Let us briefly discuss each one of these.

\begin{enumerate} 

\item Static observers

We define a static observer as an observer with four velocity given by 
\beq
u^{\mu} = \xi^{\mu}_{t}/|\xi_{t}^2|^{1/2}.
\eeq
The coordinates $r, \theta, \phi$ are constant along its worldline. Static observers cannot exist everywhere in the Kerr spacetime since $\xi^{\mu}_{t}$ is not everywhere timelike; as pointed out in the text, there is a static limit surface in which the component $g_{00}$ vanishes. The static observers cannot remain static when $r \leq r_{st}$. Instead, the dragging of inertial frames compels them to rotate with the black hole. 

\item Stationary observers

A stationary observer is an observer whose four-velocity is a combination of the two Killing vectors of the Kerr metric
\beq
u^{\mu} = (\xi^{\mu}_{t} + \Omega\,\xi^{\mu}_{\phi})/|\xi^{\mu}_{t} + \Omega\,\xi^{\mu}_{\phi}| = u^{0}(1,0,0,\Omega)
\eeq
where
$$
|\xi^{\mu}_{t} + \Omega\,\xi^{\mu}_{\phi}|^2 = - g_{\mu\nu}(\xi^{\mu}_{t} + \Omega\,\xi^{\mu}_{\phi})
(\xi^{\nu}_{t} + \Omega\,\xi^{\nu}_{\phi}),
$$
and the angular velocity of the observer is $\Omega = d\phi/dt = u^{\phi}/u^{0}$. The worldline has constant $r, \theta$. A stationary observer can exist provided $g_{\mu\nu}\,u^{\mu}\,u^{\nu} = -1$, or:
$$
\Omega^2 g_{\phi\phi} + 2\Omega g_{0\phi} + g_{00} < 0, 
$$
which implies the following requirement on the angular velocity: 
\beq
\Omega_{-} < \Omega < \Omega_{+},
\eeq
where $\Omega_{\pm} = w \pm \sqrt{w^2 - g_{00}/g_{\phi\phi}}$ and $w = - g_{0\phi}/g_{00}$. The vector $\xi^{\mu}_{t} + \Omega\,\xi^{\mu}_{\phi}$ becomes null at $r = r_{+}$ and stationary observers cannot exist inside this surface, which one identifies with the black hole's event horizon. The quantity:
\beq
\Omega_{H} = w(r_{+}) = \frac{a}{r_{+}^{2} + a^2} =  \frac{2 M a r_{+}}{(r_{+}^2 + a^2)^2}
\eeq
is then interpreted as the angular velocity of the black hole. Stationary observers just outside the horizon have an angular velocity equal to $\Omega_{H}$: they are in a state of corolation with the black hole. 

\item Zero angular momentum observer (ZAMO)

This is an observer, with timelike four-velocity $u^{\mu}$, which falls into the black hole with zero angular momentum, $u_{\phi} = 0$. This implies that at $r \to \infty$, where the metric becomes flat, one gets $u^{\phi} = 0$ (its angular velocity is zero). On the other hand, $u^{\phi} = g^{\phi 0}\,u_{0} \neq 0$ for finite $r$. The trajectory of the ZAMO has a non-zero angular velocity:
\beq
\Omega = \frac{d\phi}{dt} = \frac{d\phi/d\tau}{dt/d\tau} = \frac{u^{\phi}}{u^{0}}.
\eeq
Using the fact that $u_{\phi} = g_{\phi\phi}u^{\phi} + g_{\phi 0}u^{0} = 0$, then
\bea
\Omega &=& - \frac{g_{\phi 0}}{g_{\phi\phi}} =  \frac{2 M a r}{(r^2 + a^2)^2 - a^2\Delta\sin^2\theta} 
\nn\\
&=& \frac{\Omega_{+} + \Omega_{-}}{2}.
\eea
Notice that $\Omega \leq \Omega_{H}$. In addition, the motion of the ZAMO is corotating with the black hole. This motion takes place at constant $r, \theta$ and at constant (in time) angular velocity in $\phi$. The four velocity is given by
\beq 
u_{\mu} = -\alpha\,\delta_{\mu}^{0},\,\,\,u^{\mu} = \alpha^{-1}(1,0,0,\Omega),\,\,\,
\alpha = \sqrt{\frac{\rho^2 \Delta}{\Sigma}}.
\eeq
For nice discussions on the relevance of these observers to a couple of interesting physical problems we refer the reader the Ref.~\cite{fro2}.

\end{enumerate}

\section{Correlation functions of a massless scalar field in Kerr spacetime}
\label{B}

Before we present the correlation functions of the scalar field, we discuss briefly the quantization of the scalar field in Kerr spacetime. We will follow closely the Refs.~\cite{ote,fro3}. See also Ref.~\cite{ford}. Consider a massless scalar field with the following Lagrangian
\beq
{\cal L}(x) = \frac{1}{2}|g(x)|^{1/2}\,g^{\mu\nu}\,\partial_{\mu}\varphi\partial_{\nu}\varphi
\eeq
and hence the resulting action $S = \int d^4x {\cal L}(x)$. By setting the variation of the action with respect to $\varphi$ equal to zero, one finds the wave equation for the field. This equation is separable in the Kerr metric and the basis functions are given by
\beq
u_{\omega l m} = N_{\omega l m}\,\frac{e^{-i\omega t + i m \phi}}{(r^2 + a^2)^{1/2}}\,S_{\omega l m}(\cos\theta)\,
R_{\omega l m} (r),
\label{scalar-modes}
\eeq
where $N_{\omega l m}$ is a normalization constant, $l$ and $m$ are integers with $|m| \leq l$. $N_{\omega l m}$ is determined in such a way that the mode functions are orthonormal regarding the inner product:
$$
(u_1, u_2) = i\int_{\Sigma} d\Sigma^{\mu}\,|g_{\sigma}(x)|^{1/2}\,(u^{*}_{2}\overleftrightarrow{\partial_{\mu}} u_{1})
$$
where $d\Sigma^{\mu} = d\Sigma\,n^{\mu}$, with $n^{\mu}$ a future-directed (timelike) unit vector orthogonal to the spacelike hypersurface $\Sigma$ and $d\Sigma$ is the volume element in $\Sigma$. The function $S_{\omega l m}(x)$ is a spheroidal harmonic which satisfies
\bea
&&\biggl[\frac{d}{dx}(1 - x^2)\frac{d}{dx} - \frac{m^2}{1 - x^2} + 2ma\omega 
\nn\\
&&-\, (a\omega)^2(1 - x^2) + 
\lambda_{lm}(a\omega)\biggr]\,S_{\omega l m}(x) = 0, 
\eea
subject to regularity at $x = \pm 1$; $S_{0 l m}(x)$ is an associated Legendre function with eigenvalues $\lambda_{lm}(0) = l(l+1)$. The radial equation may be written as:
\beq
\left[\frac{d^2}{dr_{*}^2} - V_{\omega l m}(r)\right]R_{\omega l m} (r) = 0,
\eeq
where
\bea
V_{\omega l m}(r) &=& - \left(\omega - \frac{ma}{r^2 + a^2}\right)^2 + \lambda_{lm}(a\omega)\,\frac{\Delta}{(r^2 + a^2)^2} 
\nn\\
&+&\, \frac{2(Mr - a^2)\Delta}{(r^2 + a^2)^3} + \frac{3a^2\Delta^2}{(r^2 + a^2)^4}
\eea
with the tortoise coordinate $r_{*}$ defined as
\beq
r_{*} = \int\,dr\,\frac{r^2 + a^2}{\Delta} = r + \frac{1}{2\kappa_{+}}\,\ln|r - r_{+}| + \frac{1}{2\kappa_{-}}\,\ln|r - r_{-}|
\label{tortoise}
\eeq
and the surface gravity on the inner and outer horizons are given by:
\beq
\kappa_{\pm} = \frac{r_{\pm} - r_{\mp}}{2(r_{\pm}^{2} + a^2)}.
\label{surface-gravity}
\eeq
In turn, since the eigenvalues $\lambda_{lm}$ are real, this implies that $R$ and $R^{*}$ are solutions of the radial equation. 

One has the following asymptotic configurations
\bea
R^{-}_{\omega l m}(r) &\approx& e^{i\bar{\omega}r_{*}} + {\cal R}^{-}_{\omega l m}e^{-i\bar{\omega}r_{*}},\,\,\,r \to r_{+}
\nn\\
R^{-}_{\omega l m}(r) &\approx& {\cal T}^{-}_{\omega l m}e^{i\omega r_{*}},\,\,\,r \to \infty.
\label{asymp1}
\eea
and
\bea
R^{+}_{\omega l m}(r) &\approx& {\cal T}^{+}_{\omega l m}e^{-i\bar{\omega}r_{*}},\,\,\,r \to r_{+}
\nn\\
R^{+}_{\omega l m}(r) &\approx& e^{-i\omega r_{*}} + {\cal R}^{+}_{\omega l m}e^{i\omega r_{*}},\,\,\,r \to \infty.
\label{asymp2}
\eea
with $\bar{\omega} = \omega - m\Omega_{H}$. Relations between the transmission and reflection coefficients ${\cal T}^{\pm}$ and ${\cal R}^{\pm}$ reveal that $|{\cal R}^{-}|^2 > 1$ for modes with $\bar{\omega} < 0$ and $\omega > 0$. This means that they are reflected back to ${\cal H}^{+}$ with an amplitude greater than they used to have at ${\cal H}^{-}$. This classical-wave phenomenon is known as superradiance. Similar remarks apply to ${\cal R}^{+}$ at ${\cal J}^{\pm}$. The region ${\cal J}^{-}$ (${\cal J}^{+}$) is the past (future) null infinity and the region ${\cal H}^{-}$ (${\cal H}^{+}$) is the past (future) event horizon in a suitable Penrose diagram.

The existence of superradiant modes complicates the definition of positive frequency. In this case one must define states with particular properties along a given Cauchy surface. For instance, consider the Cauchy surface ${\cal J}^{-} \bigcup {\cal H}^{-}$. The past basis are then given by ($\omega > 0$):
\begin{widetext}
\bea
u^{in}_{\omega l m} &=& \left[8\pi^2\omega(r^2 + a^2)\right]^{-1/2}\,e^{-i\omega t + i m \phi}\,S_{\omega l m}(\cos\theta)\,
R^{+}_{\omega l m} (r),\,\,\,\bar{\omega} > - m\Omega_{H}
\nn\\ 
u^{up}_{\omega l m} &=& \left[8\pi^2\bar{\omega}(r^2 + a^2)\right]^{-1/2}\,e^{-i\omega t + i m \phi}\,
S_{\omega l m}(\cos\theta)\,R^{-}_{\omega l m} (r),\,\,\,\bar{\omega} > 0
\nn\\
u^{up}_{-\omega l -m} &=& \left[8\pi^2(-\bar{\omega})(r^2 + a^2)\right]^{-1/2}\,e^{i\omega t - i m \phi}\,
S_{\omega l m}(\cos\theta)\,R^{-}_{-\omega l -m} (r),\,\,\,0 >  \bar{\omega} > -m\Omega_{H},
\nn\\
\eea
where we used that $S_{-\omega l -m} = S_{\omega l m}$. Such modes are orthonormal in the sense of the standard inner product presented above, with $\Sigma = {\cal J}^{-} \bigcup {\cal H}^{-}$. This is discussed in Refs.~\cite{ote,fro3}. In addition, the future basis are given by
\bea
u^{out}_{\omega l m} &=& \left[8\pi^2\omega(r^2 + a^2)\right]^{-1/2}\,e^{-i\omega t + i m \phi}\,S_{\omega l m}(\cos\theta)\,
R^{+\,*}_{\omega l m} (r),\,\,\,\bar{\omega} > - m\Omega_{H}
\nn\\ 
u^{down}_{\omega l m} &=& \left[8\pi^2\bar{\omega}(r^2 + a^2)\right]^{-1/2}\,e^{-i\omega t + i m \phi}\,
S_{\omega l m}(\cos\theta)\,R^{-\,*}_{\omega l m} (r),\,\,\,\bar{\omega} > 0
\nn\\
u^{down}_{-\omega l -m} &=& \left[8\pi^2|\bar{\omega}|(r^2 + a^2)\right]^{-1/2}\,e^{i\omega t - i m \phi}\,
S_{\omega l m}(\cos\theta)\,R^{-\,*}_{-\omega l -m} (r),\,\,\,0 >  \bar{\omega} > -m\Omega_{H},
\nn\\
\eea
These modes are also orthonormal in the sense quoted above, with $\Sigma = {\cal J}^{+} \bigcup {\cal H}^{+}$. The asymptotic forms of all such modes can be found in Ref.~\cite{ote}.

The quantization of a scalar field in the Kerr metric may now proceed within canonical methods. As usual the canonical momentum is defined as $\Pi = |g|^{-1/2}\,\partial {\cal L}/\partial\dot{\varphi}$. We impose the canonical commutation relation:
$$
[\varphi ({\bf x}, t), \Pi({\bf x}', t)] = i\,\delta^{3}({\bf x}, {\bf x}'),
$$
where $\delta^{3}({\bf x}, {\bf x}')$ is a delta function in the hypersurface satisfying $\int d\Sigma\,\delta^{3}({\bf x}, {\bf x}') = 1$. The scalar field $\varphi$ may be expanded in terms of any of the sets of mode functions which were just presented above:
\bea
\varphi(x) &=& \sum_{l,m}\left[\int_{0}^{\infty}\,d\omega ({\hat a}^{a}_{\omega l m} u^{a}_{\omega l m} + 
{\hat a}^{a\,\dagger}_{\omega l m} u^{a*}_{\omega l m}) + \int_{\omega_{\textrm{min}}}^{\infty}\,d\omega 
({\hat a}^{b}_{\omega l m} u^{b}_{\omega l m} + {\hat a}^{b\,\dagger}_{\omega l m} u^{b*}_{\omega l m}) \right]
\nn\\
&+&\,\sum_{l,m}\int^{\omega_{\textrm{min}}}_{0}\,d\omega ({\hat a}^{b}_{-\omega l -m} u^{b}_{-\omega l -m} + 
{\hat a}^{b\,\dagger}_{-\omega l -m} u^{b*}_{-\omega l -m})
\nn\\
&=&\,\sum_{l,m}\left[\int_{0}^{\infty}\,d\omega ({\hat a}^{a}_{\omega l m} u^{a}_{\omega l m} + 
{\hat a}^{a\,\dagger}_{\omega l m} u^{a*}_{\omega l m}) + \int_{0}^{\infty}\,d\bar{\omega} 
({\hat a}^{b}_{\omega l m} u^{b}_{\omega l m} + {\hat a}^{b\,\dagger}_{\omega l m} u^{b*}_{\omega l m}) \right],
\eea
where $a = \textrm{in, out}$, $b = \textrm{up, down}$ and $\omega_{\textrm{min}} = \textrm{max}\left\{0, m\Omega_{H}\right\}$, so $\omega_{\textrm{min}} = 0$ for counter-rotating waves ($m \leq 0$) and $\omega_{\textrm{min}} = m\Omega_{H}$ for co-rotating waves ($m > 0$). Such an expansion can be inverted in order to express the creation and annihilation operators in terms of the fields and momentum operator. Hence, provided with this inverted expression, and using the canonical commutation relations as well as the inner product defined above we have the following commutation relations:
\bea
&&[{\hat a}^{a}_{\omega l m}, {\hat a}^{a\,\dagger}_{\omega' l' m'}] = \delta(\omega - \omega')\,\delta_{ll'}\,\delta_{mm'},\,\,\,
\bar{\omega} > - m\Omega_{H}
\nn\\
&&[{\hat a}^{b}_{\omega l m}, {\hat a}^{b\,\dagger}_{\omega' l' m'}] = \delta(\omega - \omega')\,\delta_{ll'}\,\delta_{mm'},
\,\,\,\bar{\omega} > 0
\nn\\
&&[{\hat a}^{b}_{-\omega l -m}, {\hat a}^{b\,\dagger}_{-\omega' l' -m'}] = \delta(\omega - \omega')\,\delta_{ll'}\,\delta_{mm'},
\,\,\, 0 >  \bar{\omega} > -m\Omega_{H}
\nn\\
\eea
with all other commutators vanishing. We define a past Boulware vacuum state by
\bea
&& {\hat a}^{in}_{\omega l m}|B^{-}\rangle = 0,\,\,\,\bar{\omega} > - m\Omega_{H}
\nn\\
&& {\hat a}^{up}_{\omega l m}|B^{-}\rangle = 0,\,\,\,\bar{\omega} > 0
\nn\\
&& {\hat a}^{up}_{-\omega l -m}|B^{-}\rangle = 0,\,\,\, 0 >  \bar{\omega} > -m\Omega_{H},
\eea
corresponding to an absence of particles from ${\cal J}^{-}$ and ${\cal H}^{-}$. In the same way, the future Boulware vacuum state is defined by
\bea
&& {\hat a}^{out}_{\omega l m}|B^{+}\rangle = 0,\,\,\,\bar{\omega} > - m\Omega_{H}
\nn\\
&& {\hat a}^{down}_{\omega l m}|B^{+}\rangle = 0,\,\,\,\bar{\omega} > 0
\nn\\
&& {\hat a}^{down}_{-\omega l -m}|B^{+}\rangle = 0,\,\,\, 0 >  \bar{\omega} > -m\Omega_{H},
\eea
corresponding to an absence of particles from ${\cal J}^{+}$ and ${\cal H}^{+}$. 

The past Unruh state $|U^{-}\rangle$ is easily defined as that state empty at ${\cal J}^{-}$ but with the up modes (natural modes on ${\cal H}^{-}$) thermally populated. One may also define the future Unruh state $|U^{+}\rangle$ as that state empty at ${\cal J}^{+}$ but with the down modes (natural modes on ${\cal H}^{+}$) thermally populated. However, it is $|U^{-}\rangle$ that mimics the state arising at late times from the collapse of a star to a black hole. Hence this is the Unruh vacuum state that we shall adopt in this work. On the other hand, as mentioned above, in the absence of such a true Hartle-Hawking vacuum we consider two candidates in order to define a thermal state with most, but not all, of the properties of the Hartle-Hawking state: The vacuum state 
$|CCH\rangle$ was introduced by Candelas, Chrzanowski and Howard~\cite{candelas2}, which is constructed by thermalizing the in and up modes with respect to their natural energy. This definition gives a state which does not respect the simultaneous $t-\phi$ reversal invariance of Kerr spacetime. By contrast, Frolov and Thorne in~\cite{fro3} have established a vacuum state $|FT\rangle$ which is formally invariant under simultaneous $t-\phi$ reversal. For a detailed discussion on these three vacuum states discussed in this paragraph, we refer the reader the Ref.~\cite{fro3}. 

Finally we are in position to determine all the relevant correlation functions. One has that:
\bea
D_{B^{\pm}}(x,x') &=& \langle B^{\pm} |\{\varphi(x),\varphi(x')\}| B^{\pm} \rangle
\nn\\
&=&\, \sum_{l,m}\left\{\int_{0}^{\infty}\,d\omega \left[u^{a}_{\omega l m}(x)u^{a*}_{\omega l m}(x') 
+  u^{a}_{\omega l m}(x')u^{a*}_{\omega l m}(x)\right]\right\} 
\nn\\
&+&\, \sum_{l,m}\left\{\int_{0}^{\infty}\,d\bar{\omega} \left[u^{b}_{\omega l m}(x) u^{b*}_{\omega l m}(x')
+u^{b}_{\omega l m}(x') u^{b*}_{\omega l m}(x)\right] \right\},
\label{hada-boul}
\eea
for the Hadamard's elementary function of the Boulware vacuum states, whereas the associated Pauli-Jordan function reads
\bea
\Delta_{B^{\pm}}(x,x') &=& \langle B^{\pm} |[\varphi(x),\varphi(x')]| B^{\pm} \rangle
\nn\\
&=&\, \sum_{l,m}\left\{\int_{0}^{\infty}\,d\omega \left[u^{a}_{\omega l m}(x)u^{a*}_{\omega l m}(x') 
-  u^{a}_{\omega l m}(x')u^{a*}_{\omega l m}(x)\right]\right\} 
\nn\\
&+&\, \sum_{l,m}\left\{\int_{0}^{\infty}\,d\bar{\omega} \left[u^{b}_{\omega l m}(x) u^{b*}_{\omega l m}(x')
- u^{b}_{\omega l m}(x') u^{b*}_{\omega l m}(x)\right] \right\}.
\label{pauli-boul}
\eea
Likewise the Hadamard's elementary function of the Unruh vacuum state is given by:
\bea
D_{U^{-}}(x,x') &=& \langle U^{-} |\{\varphi(x),\varphi(x')\}| U^{-} \rangle
\nn\\
&=&\, \sum_{l,m}\left\{\int_{0}^{\infty}\,d\omega \left[u^{in}_{\omega l m}(x)u^{in\,*}_{\omega l m}(x') 
+  u^{in}_{\omega l m}(x')u^{in\,*}_{\omega l m}(x)\right]\right\} 
\nn\\
&+&\, \sum_{l,m}\left\{\int_{0}^{\infty}\,d\bar{\omega} \coth\left( \frac{\pi\bar{\omega}}{\kappa_{+}} \right) \left[u^{up}_{\omega l m}(x) u^{up\,*}_{\omega l m}(x')
+u^{up}_{\omega l m}(x') u^{up\,*}_{\omega l m}(x)\right] \right\},
\label{hada-unruh}
\eea
whereas the associated Pauli-Jordan function reads
\bea
\Delta_{U^{-}}(x,x') &=& \langle U^{-} |[\varphi(x),\varphi(x')]| U^{-} \rangle
\nn\\
&=&\, \sum_{l,m}\left\{\int_{0}^{\infty}\,d\omega \left[u^{in}_{\omega l m}(x)u^{in\,*}_{\omega l m}(x') 
-  u^{in}_{\omega l m}(x')u^{in\,*}_{\omega l m}(x)\right]\right\} 
\nn\\
&+&\, \sum_{l,m}\left\{\int_{0}^{\infty}\,d\bar{\omega} \left[u^{up}_{\omega l m}(x) u^{up\,*}_{\omega l m}(x')
- u^{up}_{\omega l m}(x') u^{up\,*}_{\omega l m}(x)\right] \right\}.
\label{pauli-unruh}
\eea
In turn, the Hadamard's elementary function with respect to the Candelas-Chrzanowski-Howard vacuum state is given by:
\bea
D_{CCH}(x,x') &=& \langle CCH |\{\varphi(x),\varphi(x')\}| CCH \rangle
\nn\\
&=&\, \sum_{l,m}\left\{\int_{0}^{\infty}\,d\omega\coth\left( \frac{\pi\omega}{\kappa_{+}} \right)
 \left[u^{in}_{\omega l m}(x)u^{in\,*}_{\omega l m}(x') 
+  u^{in}_{\omega l m}(x')u^{in\,*}_{\omega l m}(x)\right]\right\} 
\nn\\
&+&\, \sum_{l,m}\left\{\int_{0}^{\infty}\,d\bar{\omega} \coth\left( \frac{\pi\bar{\omega}}{\kappa_{+}} \right) \left[u^{up}_{\omega l m}(x) u^{up\,*}_{\omega l m}(x')
+u^{up}_{\omega l m}(x') u^{up\,*}_{\omega l m}(x)\right] \right\}.
\label{hada-cCh}
\eea
The associated Pauli-Jordan function reads
\bea
\Delta_{CCH}(x,x') &=& \langle CCH |[\varphi(x),\varphi(x')]| CCH \rangle
\nn\\
&=&\, \sum_{l,m}\left\{\int_{0}^{\infty}\,d\omega \left[u^{in}_{\omega l m}(x)u^{in\,*}_{\omega l m}(x') 
-  u^{in}_{\omega l m}(x')u^{in\,*}_{\omega l m}(x)\right]\right\} 
\nn\\
&+&\, \sum_{l,m}\left\{\int_{0}^{\infty}\,d\bar{\omega} \left[u^{up}_{\omega l m}(x) u^{up\,*}_{\omega l m}(x')
- u^{up}_{\omega l m}(x') u^{up\,*}_{\omega l m}(x)\right] \right\}.
\label{pauli-cCh}
\eea
At last, the Hadamard's elementary function associated with the Frolov-Thorne vacuum state is given by:
\bea
D_{FT}(x,x') &=& \langle FT |{\hat \eta}\varphi(x) {\hat \eta} \varphi(x') {\hat \eta}| FT \rangle +
\langle FT |{\hat \eta}\varphi(x') {\hat \eta} \varphi(x) {\hat \eta}| FT \rangle
\nn\\
&=&\, \sum_{l,m}\left\{\int_{0}^{\infty}\,d\omega\coth\left( \frac{\pi\bar{\omega}}{\kappa_{+}} \right)
 \left[u^{in}_{\omega l m}(x)u^{in\,*}_{\omega l m}(x') 
+  u^{in}_{\omega l m}(x')u^{in\,*}_{\omega l m}(x)\right]\right\} 
\nn\\
&+&\, \sum_{l,m}\left\{\int_{0}^{\infty}\,d\bar{\omega} \coth\left( \frac{\pi\bar{\omega}}{\kappa_{+}} \right) \left[u^{up}_{\omega l m}(x) u^{up\,*}_{\omega l m}(x')
+u^{up}_{\omega l m}(x') u^{up\,*}_{\omega l m}(x)\right] \right\}.
\label{hada-frolov}
\eea
where the number operator ${\hat \eta}^2 = 1$, see Ref.~\cite{fro3}. The Pauli-Jordan function of the Frolov-Thorne vacuum state reads:
\bea
\Delta_{FT}(x,x') &=& \langle FT |{\hat \eta}\varphi(x) {\hat \eta} \varphi(x') {\hat \eta}| FT \rangle -
\langle FT |{\hat \eta}\varphi(x') {\hat \eta} \varphi(x) {\hat \eta}| FT \rangle
\nn\\
&=&\, \sum_{l,m}\left\{\int_{0}^{\infty}\,d\omega \left[u^{in}_{\omega l m}(x)u^{in\,*}_{\omega l m}(x') 
-  u^{in}_{\omega l m}(x')u^{in\,*}_{\omega l m}(x)\right]\right\} 
\nn\\
&+&\, \sum_{l,m}\left\{\int_{0}^{\infty}\,d\bar{\omega} \left[u^{up}_{\omega l m}(x) u^{up\,*}_{\omega l m}(x')
- u^{up}_{\omega l m}(x') u^{up\,*}_{\omega l m}(x)\right] \right\}.
\label{pauli-frolov}
\eea

\end{widetext}


\begin{thebibliography}{99}
%

\bibitem{frolov}
	V. P. Frolov and I. D. Novikov,
	{\it Black Hole Physics: Basic Concepts and New Developments}
	(Kluwer Academic Publishers, Netherlands,1998).

\bibitem{ellis}
	S. W. Hawking and G. F. R. Ellis,
	{\it The Large Scale Structure of Space-Time}
	(Cambridge University Press, Cambridge, 1975).

\bibitem{cardoso}
	V. Cardoso and P. Pani,
 	Class. Quantum Grav. {\bf 30}, 045011 (2013).

\bibitem{rot1}
 	E. M. Butterworth and J. R. Ipser, 
  	Astrophys. J. {\bf 204}, 200 (1976).

\bibitem{rot2}
	B. F. Schutz and N. Comins,
	Mon. Not. Roy. Astron. Soc. {\bf 182}, 69 (1978).

\bibitem{zeldovich}
	Ya. B. Zel'dovich,
	Sov. Phys. JETP Lett. {\bf 14}, 180 (1971).

\bibitem{misner}
	C. W. Misner, 
 	Phys. Rev. Lett. {\bf 28}, 994 (1972).

\bibitem{teu3}
	W. H. Press and S.A. Teukolsky,
	Nature (London) {\bf 238}, 211 (1972).

\bibitem{dub}
	A. Arvanitaki and S. Dubovsky,
	Phys. Rev. D {\bf 83}, 044026 (2011).

\bibitem{pani}
	P. Pani, V. Cardoso, L. Gualtieri, E. Berti and A. Ishibashi,
	Phys. Rev. Lett. {\bf 109}, 131102 (2012).

\bibitem{brito}
	R. Brito, V. Cardoso and P. Pani,
	Phys. Rev. D. {\bf 88}, 023514 (2013).

\bibitem{review}
	R. Brito, V. Cardoso and P. Pani,
	{\it Superradiance: Energy Extraction, Black-Hole Bombs and Implications for Astrophysics and Particle Physics}
	(Springer, London, 2015).

\bibitem{meier}
	D. L. Meier, 
	New Astron. Rev. {\bf 47}, 667 (2003).

\bibitem{mirabel}
	I. F. Mirabel and L. F. Rodr\'iguez,
	Annu. Rev. Astron. Astrophys. {\bf 37}, 409 (1999).

\bibitem{blandford}
	R. Blandford, E. Agol, A. Broderick, J. Heyl, L. Koopmans, and H. W. Lee, in
	{\it Astrophysical Spectropolarimetry (Proceedings of the XII Canary Islands Winter School of Astrophysics)},
	edited by J. Trujillo-Bueno, F. Moreno-Insertis and F. S\'anchez
	(Cambridge University Press, Cambridge, 2002),
	Vol.1, 177.

\bibitem{Lynden}
	D. Lynden-Bell,
	Nature {\bf 223}, 690 (1969).

\bibitem{blandford2}
	R. Blandford and R. Znajek,
	Mon. Not. Roy. Astron. Soc. {\bf 179}, 433 (1977).

\bibitem{penrose}
	R. Penrose and R. M. Floyd, 
	Nature {\bf 229}, 177 (1971).

\bibitem{kom}
	S. Komissarov,
	J. Korean Phys. Soc. {\bf 54}, 2503 (2009).

\bibitem{lasota}
	J. P. Lasota, E. Gourgoulhon, M. Abramowicz, A. Tchekhovskoy and R. Narayan,
	Phys. Rev. D {\bf 89}, 024041 (2014).

\bibitem{penner}
	M. Richartz, S. Weinfurtner, A. J. Penner and W. G. Unruh,
	Phys. Rev. D {\bf 80}, 124016 (2009).

\bibitem{sta}
	A. A. Starobinskii,
	Zh. Eksp. Teor. Fiz. {\bf 64}, 48 (1973).

\bibitem{unruh-sta}
	W. G. Unruh,
	Phys. Rev. D {\bf 10}, 3194 (1974).

\bibitem{davies}
	A. L. Matacz, P. C. W. Davies and A. C. Ottewill,
	Phys. Rev. D {\bf 47}, 1557 (1993).

\bibitem{haw1}
	S. W. Hawking,
	Commun. Math. Phys. {\bf 43}, 199 (1975).

\bibitem{mil1}
	P. W. Milonni, 
	Phys. Rep. {\bf 25}, 1 (1976).

\bibitem{cohen2} 
	J. Dalibard, J. Dupont-Roc, and C. Cohen-Tannoudji, 
	J. Phys. (Paris) {\bf 43}, 1617 (1982).

\bibitem{cohen3} 
	J. Dalibard, J. Dupont-Roc, and C. Cohen-Tannoudji, 
	J. Phys. (Paris) {\bf 45}, 637 (1984).


\bibitem{aud1} 
	J. Audretsch and R. M\"uller, 
	Phys. Rev. A {\bf 50}, 1755 (1994).

\bibitem{aud2} 
	J. Audretsch and R. M\"uller, 
	Class. Quant. Grav. {\bf 12}, 2927 (1995).

\bibitem{china5} 
	H. Yu and W. Zhou,  
	Phys. Rev. D {\bf 76}, 044023 (2007).

\bibitem{china6} 
	H. Yu and W. Zhou, 
	Phys. Rev. D {\bf 76} 027503 (2007).

\bibitem{china} 
	W. Zhou and H. Yu, 
	Class. Quantum Grav. {\bf 29}, 085003 (2012).

\bibitem{china1} 
	Z. Zhu, H. Yu and S. Lu,
	 Phys. Rev. D {\bf 73}, 107501 (2006).

\bibitem{china2} 
	Z. Zhu and H. Yu, 
	Phys. Rev. A {\bf 82}, 042108 (2010).

\bibitem{rizz} 
	L. Rizzuto and S. Spagnolo, 
	Phys. Scr. {\bf T143} 014021 (2011).

\bibitem{ng1} 
	G. Menezes and N. F. Svaiter,  
	Phys. Rev. A {\bf 92}, 062131 (2015). 

\bibitem{ng2} 
	G. Menezes and N. F. Svaiter, 
	Phys. Rev. A {\bf 93}, 052117 (2016).

\bibitem{ng3}
	G. Menezes,
	Phys. Rev. D {\bf 94}, 105008 (2016). 

\bibitem{ref1}
	L. Rizzuto, M. Lattuca, J. Marino, A. Noto, S. Spagnolo, W. Zhou and R. Passante
	Phys. Rev. A {\bf 94}, 012121 (2016).

\bibitem{ref2}
	W. Zhou, R. Passante and L. Rizzuto
	Phys. Rev. D {\bf 94}, 105025 (2016).

\bibitem{birrel}
	N. D. Birrell and P. C. W. Davies,
	{\it Quantum Fields in Curved Space}
	(Cambridge University Press, New York, 1982).

\bibitem{sciama}
	D. W. Sciama, P. Candelas and D. Deutsch,
	Adv. Phys {\bf 30}, 327 (1981).

\bibitem{unruh} 
	W. G. Unruh,  
	Phys. Rev. D {\bf 14}, 870 (1976).

\bibitem{dw2}
	B. S. DeWitt, in
	{\it General Relativity: An Einstein Centenary Survey},
	edited by S. W. Hawking and W. Israel,
	(Cambridge University Press, Cambridge, 1979),
	680.

\bibitem{ote}
	A. C. Ottewill and E. Winstanley,
	Phys. Rev. D {\bf 62}, 084018 (2000).

\bibitem{boul}
	D. G. Boulware
	Phys. Rev. D {\bf 11}, 1404 (1975).

\bibitem{wald}
	B. S. Kay and R. M. Wald,
	Phys. Rep. {\bf 207}, 49 (1991).

\bibitem{candelas2}
	P. Candelas, P. Chrzanowski and K. W. Howard
	Phys. Rev. D {\bf 24}, 297 (1981).

\bibitem{fro3}
	V. P. Frolov and K. S. Thorne, 
	Phys. Rev. D {\bf 39}, 2125 (1989).

\bibitem{rot3}
	P. C. W. Davies, T. Dray and C. A. Manogue, 
	Phys. Rev. D {\bf 53},4382 (1996).

\bibitem{rot4}
	V. A. De Lorenci and N. F. Svaiter,
	Found. Phys. {\bf 29}, 1233 (1999).

\bibitem{xian}
	X.-M. Liu and W.-B. Liu,
	Adv. High Energy Physics {\bf 2014}, 794626 (2014).

\bibitem{cardoso2}
	V. Cardoso and O. J. Dias, 
	Phys. Rev. D {\bf 70}, 084011 (2004).

\bibitem{dias}
	O. J. Dias, P. Figueras, S. Minwalla, P. Mitra, R. Monteiro and J. E. Santos, 
	JHEP {\bf 1208}, 117 (2012).

\bibitem{dias2}
	O. J. Dias, G. T. Horowitz and J. E. Santos, 
	JHEP {\bf 1107}, 115 (2011).

\bibitem{koba}
	T. Kobayashi and A. Tomimatsu,
	Phys. Rev. D {\bf 82}, 084026 (2010).

\bibitem{peres}
	A. Peres and D. R. Terno, 
	Rev. Mod. Phys. {\bf 76}, 93 (2004).

\bibitem{cqg}
	Relativistic quantum information, 
	special issue of Classical and Quantum Gravity {\bf 29}, 11 (2012).

\bibitem{fuentes}
	P. M. Alsing and I. Fuentes, I.,
	Class. Quantum Grav. {\bf 29}, 224001 (2012).

\bibitem{marino}
	J. Marino, A. Noto and R. Passante,
	Phys. Rev. Lett. {\bf 113}, 020403 (2014).

\bibitem{poisson}
	E. Poisson,
	{\it A Relativist's Toolkit: The mathematics of black-hole mechanics}
	(Cambridge University Press, Cambridge, 2007).

\bibitem{fro2}
	A. V. Frolov and V. P. Frolov, 
	Phys. Rev. D {\bf 90}, 124010 (2014).

\bibitem{ford}
	L. H. Ford,
	Phys. Rev. D {\bf 12}, 2963 (1975).


\end{thebibliography}
\end{document}